\documentclass[12pt,3p,times]{elsarticle}
\usepackage{color,graphicx}
\usepackage{color,graphicx,amsmath,amssymb}

\def\lvec#1{\setbox0=\hbox{$#1$}
    \setbox1=\hbox{$\scriptstyle\leftarrow$}
    #1\kern-\wd0\smash{
    \raise\ht0\hbox{$\raise1pt\hbox{$\scriptstyle\leftarrow$}$}}
    \kern-\wd1\kern\wd0}
\def\rvec#1{\setbox0=\hbox{$#1$}
    \setbox1=\hbox{$\scriptstyle\rightarrow$}
    #1\kern-\wd0\smash{
    \raise\ht0\hbox{$\raise1pt\hbox{$\scriptstyle\rightarrow$}$}}
    \kern-\wd1\kern\wd0}
\def\diracstar#1#2{
    \setbox0=\hbox{$\gamma$}\setbox1=\hbox{$\gamma_{#1}$}\includegraphics[]{QNEW_36-eps-converted-to.pdf}

    \gamma_{#1}\kern-\wd1\kern\wd0
    \smash{\raise4.5pt\hbox{$\scriptstyle#2$}}}
\def\tr{\,\hbox{tr}\,}

\newcommand{\beq}{\begin{equation}}
\newcommand{\eeq}{\end{equation}}
\newcommand{\beqn}{\begin{eqnarray}}
\newcommand{\eeqn}{\end{eqnarray}}
\newcommand{\nn}{\nonumber}

\renewcommand\appendix{\par
  \setcounter{section}{0}%
  \setcounter{subsection}{0}%
  \setcounter{equation}{0}
 \gdef\thefigure{\@Alph\c@section.\arabic{figure}}%
 \gdef\thetable{\@Alph\c@section.\arabic{table}}%
 \gdef\thesection{\appendixname~\@Alph\c@section}\@addtoreset{equation}{section}%
\gdef\theequation{\@Alph\c@section.\arabic{equation}}%
  \addtocontents{toc}{\string\let\string\numberline\string\tmptocnumberline}{}{}
}

\usepackage{chir_layout}

\begin{document}
\begin{frontmatter}
%%%%%%%%%%%%%%%%%%%%%%%%%
%\begin{titlepage}\pagestyle{empty}\date{}
\title{{
\bf A road to an elementary particle physics model \\ with no Higgs~-~II }
\vspace*{-4mm}}

\author{G.C.\ Rossi}

\address{$^{a}$ {\small Dipartimento di Fisica, Universit\`a di  Roma
  ``{\it Tor Vergata}'' \\ INFN, Sezione di Roma  ``{\it Tor Vergata}'' }\\
  {\small Via della Ricerca Scientifica - 00133 Roma, Italy}\\
  {\small Centro Fermi - Museo Storico della Fisica e Centro Studi e Ricerche E.\ Fermi, \\ Via Panisperna 89a, 00184 Roma, Italy}}

%%%%%%%%%%%%%%%%%%%%%%%%%%%%%%%%%%%%%%

\vspace*{-3mm}
\begin{abstract}
This is the second of two companion papers in which we continue developing the construction of an elementary particle model with no Higgs. Here we show that the recently identified non-perturbative field-theoretical feature, alternative to the Higgs mechanism and capable of giving masses to quarks, Tera-quarks and $W$, can also provide mass to leptons and Tera-leptons when the model is extended to include, besides strong, Tera-strong and weak interactions, also hypercharge. In the present approach elementary fermion masses are not free parameters but are determined by the dynamics of the theory. We derive parametric formulae for elementary particle masses from which we can ``predict'' the order of magnitude of the scale of the new Tera-interaction and get crude numerical estimates for mass ratios in fair agreement with phenomenology. The interest of considering elementary particle models endowed with this kind of non-perturbative mass generation mechanism is that they allow solving some of the conceptual problems of the present formulation of the Standard Model, namely origin of the electroweak scale and naturalness.
\end{abstract}
\end{frontmatter}

\section{Introduction}
\label{sec:INTRO}

This investigation is a continuation of the work initiated in ref.~\cite{Frezzotti:2014wja} (see also~\cite{Frezzotti:2013raa}) and expanded with the introduction of weak and Tera-interactions in ref.~\cite{Rossi:2023xhl}~\footnote{In the following we will refer to this paper as~(I).}, where it was proved that the non-perturbative (NP) mechanism found in~\cite{Frezzotti:2014wja} to be able to give mass to elementary fermions (as confirmed by explicit lattice simulations in~\cite{Capitani:2019syo}), can provide mass also to $W$ and Tera-quarks~\footnote{Preliminary results in this direction were presented in refs.~\cite{Frezzotti:2018zsy,Frezzotti:2019npa,Rossi:2022vlr,Rossi:2022xpa,Rossi:2023jjv}.}. 

As a further step towards the construction of a realistic elementary particle theory, in this paper we extend the model by including hypercharge interactions and introducing leptons and Tera-leptons (i.e.\ particles that are singlets under SU($N_c=3$) color gauge interactions). We shall see that leptons as well as Tera-leptons acquire a NP mass by the same dynamical mechanism by which quarks, Tera-quarks and $W$'s do.\ This mechanism can thus be used as a viable alternative to the Higgs scenario in the construction of a phenomenologically acceptable beyond-the-Standard-Model model (bSMm). 

Models of the kind advocated in refs.~\cite{Frezzotti:2014wja} and~(I) (and extended here) are rather interesting as they have a number of appealing theoretical features that allow solving some of the problems of the present formulation of the Standard Model (SM), as we briefly detail below.

1) There isn't anymore a Higgs mass tuning problem~\cite{Susskind:1978ms,THOOFT} as there is no fundamental Higgs around.

2) Elementary particle masses are not unconstrained parameters of the Lagrangian, but they are determined by the dynamics of the theory.

3) Since masses are naturally proportional to the RGI scale of the theory, to cope with the phenomenological values of the top-quark and weak bosons masses, the RGI scale of the underlying fundamental theory must be much larger than $\Lambda_{QCD}$ and O(a few TeV's). This implies that there must exist a yet unobserved sector of massive fermions (Tera-fermions) subjected, besides Standard Model interactions, to some kind of super-strong gauge interactions (Tera-interactions) so that the full theory (including Standard Model and Tera-particles) will have an RGI scale $\Lambda_{RGI}$ (we call it $\Lambda_T$, where $T$ stands for Tera) $\gg \Lambda_{QCD}$ lying in the few TeV region.  

4) The electroweak (EW) scale is naturally interpreted as (a fraction of) the RGI scale of the theory which is to be identified with the Tera-strong $\Lambda_T$ parameter. 

5) It was proven in~\cite{Frezzotti:2016bes} that with a reasonable choice of the spectrum and hypercharges of Tera-particles a model extending the SM degrees of freedom (dof's) with the inclusion of the Tera-sector leads to a theory with unification of the running of the U(1)$_Y$, SU(2)$_L$ and SU($N_c=3$) gauge couplings at a scale of about $10^{18}$~GeV. This somewhat large unification scale may be welcome in view of the present proton life-time bound $\tau_{\rm prot} > 1.7 \times10^{34}$ years~\cite{Bajc:2016qcc}.

6) We show that our mass formulae allow to ``predict'' the order of magnitude of the scale $\Lambda_T$ and the masses of the heaviest family in terms of a fundamental energy scale that for reasons explained in the text we take to be the $W$ mass.

As a final observation we need to recall that, like we show in~(I), the low energy effective Lagrangian (LEEL) of the model valid for (momenta)$^2\ll \Lambda_T^2$ looks very much like the SM Lagrangian. Indeed, in our scheme the125~GeV resonance detected at LHC~\cite{Aad:2012tfa,Chatrchyan:2012xdj}, that suggestively we denote by $h$, is interpreted as a composite $W^+W^-/ZZ$ scalar particle bound by Tera-strong exchanges, left behind after integrating out the heavy Tera-dof's. Since its mass is very small compared to the conjectured value of the $\Lambda_T$ scale, it needs to be accounted for in describing the low energy dynamics of the model. Ignoring perturbative corrections (see sect.~4 of~(I)), differences between the $d=4$ part of the LEEL of the model and the SM Lagrangian may only appear in the trilinear and quadrilinear couplings of the $h$ self-interactions.

\subsection{Plan of the paper}

The plan of the paper is as follows. In sect.~\ref{sec:INHYP} we provide the full expression of the Lagrangian of a model where besides quarks, Tera-quarks and $W$'s with their strong, Tera-strong and weak interactions, hypercharge interactions are switched on and leptons and Tera-leptons are introduced. In sect.~\ref{sec:MASS}, extending the analysis and the results of refs.~\cite{Frezzotti:2014wja} and~(I), we provide the leading loop order parametric expression of the masses of all the elementary particles of the model as they emerge in the Nambu--Goldstone (NG) phase of the critical 
theory. In sect.~\ref{sec:RDU} we discuss the issue of universality of the mass formulae and the related question of the predictive power of the present  approach. In sect.~\ref{sec:SPCS} we show how our mass formulae allow to ``predict'' the order of magnitude of $\Lambda_T$ and derive crude numerical estimates for the fermion masses of the heaviest family in terms of the $W$ mass, in fair agreement with phenomenology. In sect.~\ref{sec:QEL} we give the expression of the $d\leq 4$ part of the Quantum Effective Lagrangian (QEL) that describes the physics of the model~\footnote{By Quantum Effective Lagrangian we mean the generating functional of the 1PI vertices of the theory from which the full quantum information of the model can be directly extracted.}. In sect.~\ref{sec:HIGGS} we provide some support for our interpretation of the 125~GeV state identified at LHC, as a $W^+W^-/ZZ$ composite state bound by Tera-particle exchanges, based on the results of an oversimplified Bethe--Salpeter-like approach. Conclusions and an outlook of our future lines of investigation can be found in sect.~\ref{sec:CONC}. In Appendix~\ref{sec:APPA} for completeness we recall the hypercharge assignments of SM and of Tera-particles yielding U(1)$_Y$, SU(2)$_L$ and SU($N_c=3$) gauge coupling unification~\cite{Frezzotti:2016bes}. In Appendix~\ref{sec:APPB} we provide some understanding of the microscopic mechanism responsible for the emergence of the $d=6$ NP O($b^2$) Symanzik operators that are at the basis of the dynamical mass generation mechanism. In Appendix~\ref{sec:APPC} we discuss the $\rho$ dependence of physical quantities and masses in particular. In Appendix~\ref{sec:APPD} we compute in the non-relativistic approximation the $WW$ binding energy due to exchanges of Tera-particles.

\section{Introducing hypercharge, leptons and Tera-leptons}
\label{sec:INHYP}

The Lagrangian reported in equations from~(3.1) to~(3.5) of ref.~(I) can be straightforwardly extended to introduce hypercharge interactions and include leptons and Tera-leptons. For this it is enough to add appropriate kinetic, Yukawa and Wilson-like terms for the new particles and suitably extend the expression of the covariant derivatives. In this paper we restrict ourselves to the one family case and drop the flavour index because, lacking for the moment a mechanism to remove family degeneracy, appending a flavour index to quarks and leptons would not add anything useful while making notations more clumsy. Even restricting to the one family case and in the limit of unbroken weak isospin, one finds the rather lengthy expression
\beqn
\hspace{-1.6cm}&&{\cal L}(q,\ell,Q,L;\Phi;A,G,W,B)={\cal L}_{kin}(q,\ell,Q,L;\Phi;A,G,W,B)+{\cal V}(\Phi)+\nn\\
\hspace{-1.6cm}&&\qquad+{{\cal L}_{Yuk}}(q,\ell,Q,L;\Phi) +{{\cal L}_{Wil}}(q,\ell,Q,L;\Phi;A,G,W,B) \label{LTOT}
\eeqn
\beqn
\hspace{-.8cm}&&\bullet\,\,{\cal L}_{kin}(q,\ell,Q,L;\Phi;A,G,W,B)= \nn\\
\hspace{-.8cm}&&\hspace{.5cm}=\frac{1}{4}\Big{(}F^A\cdot F^A+F^G\cdot F^G+F^W\cdot F^W+F^B\cdot F^B\Big{)}+\nn\\
\hspace{-.8cm}&& \hspace{.5cm}+\Big{[}\bar q_L\,\, /\!\!\!\! {\cal D}^{ABW} q_L+\bar q_R\,\, /\!\!\!\!  {\cal D}^{AB} q_R +\bar \ell_L \,\, /\!\!\!\!  {\cal D}^{BW} \ell_L+\bar \ell_R \,\, /\!\!\!\!  {\cal D}^{B} \ell_R\Big{]} +\nn\\
\hspace{-.8cm}&&\hspace{.5cm}+\Big{[}\bar Q_L \,\, /\!\!\!\!  {\cal D}^{AGBW} Q_L+\bar Q_R \,\, /\!\!\!\!  {\cal D}^{AGB} Q_R+\bar L_L\,\, /\!\!\!\!  {\cal D}^{GBW} L_L+\bar L_R \,\, /\!\!\!\!  {\cal D}^{GB} L_R \Big{]} +\nn\\
\hspace{-.8cm}&&\hspace{.5cm}+\frac{k_b}{2}{\tr}\big{[}({\cal D}\,^{WB}_\mu\Phi)^\dagger  {\cal D}^{BW}_\mu\Phi\big{]}\label{LK}\\
\hspace{-.8cm}&&\bullet\,\,{\cal V}(\Phi)= \frac{\mu_0^2}{2}k_b{\tr}\big{[}\Phi^\dagger\Phi\big{]}+\frac{\lambda_0}{4}\big{(}k_b{\tr}\big{[}\Phi^\dagger\Phi\big{]}\big{)}^2 \label{LV}\\
\hspace{-.8cm}&&\bullet\,\,{{\cal L}_{Yuk}}(q,\ell,Q,L;\Phi)=\sum_{f=q,\ell,Q,L}{\eta_f}\,\big{(} \bar f_L\Phi\, f_R+{\mbox{hc}}\big{)}   \label{LY}\\ 
\hspace{-.8cm}&&\bullet\,\,{{\cal L}_{Wil}}(q,\ell,Q,L;\Phi;A,G,W,B)= \nn\\
\hspace{-.8cm}&&\hspace{.5cm}=\frac{{b^2}}{2}{\rho_q}\,\big{(}\bar q_L{\overleftarrow {\cal D}}\,^{ABW}_\mu\Phi {\cal D}^{AB}_\mu q_R+{\mbox{hc}}\big{)} +\frac{{b^2}}{2}{\rho_\ell}\,\big{(}\bar \ell_L{\overleftarrow {\cal D}}\,^{BW}_\mu\Phi {\cal D}^{B}_\mu \ell_R+{\mbox{hc}}\big{)} +\nn\\
\hspace{-.8cm}&&\hspace{.5cm}+\frac{{b^2}}{2}{\rho_Q}\,\big{(}\bar Q_L{\overleftarrow {\cal D}}\,^{AGBW}_\mu\Phi {\cal D}^{AGB}_\mu Q_R\!+\!{\mbox{hc}}\big{)}  \!+\frac{{b^2}}{2}{\rho_L}\,\big{(}\bar L_L{\overleftarrow {\cal D}}\,^{GBW}_\mu\Phi {\cal D}^{GB}_\mu L_R+{\mbox{hc}}\big{)} \label{LW}\, .
\eeqn
Following the notations of refs.~\cite{Frezzotti:2014wja} and~(I), above we have indicated by ${\cal D}_\mu^X$ the covariant derivative with respect to the group of transformations of which $\{X\}$ are the associated gauge bosons. The most general expression of the covariant derivative is
\begin{equation}
D_\mu^{AGBW}=\partial_\mu-iYg_YB_\mu-ig_w\tau^r W^r_\mu-ig_s\frac{\lambda^a}{2} A^a_\mu-ig_T\frac{\lambda^{\alpha}_T}{2} G^{\alpha}_\mu \, , \label{DER}
\end{equation}
where $Y, \tau^r (r=1,2,3), \lambda^a (a=1,2, \ldots, N_c^2-1)$ and $\lambda^{\alpha}_T\, (\alpha=1,2,\ldots, N_T^2-1)$ are, respectively, the U(1)$_Y$ hypercharge and the generators of the SU(2)$_L$, SU($N_c=3$) and SU($N_T=3$) group with $g_Y, g_w, g_s, g_T$ denoting the corresponding gauge couplings~\footnote{There should be no confusion between the subscript $L$ indicating left chirality and the short-hand $L$ by which we denote Tera-leptons.}~\footnote{Following~\cite{PESK}, in our hypercharge normalization we have $Q=T_3+Y$, see Tables~\ref{tab:standard} and~\ref{tab:alternative} in Appendix~\ref{sec:APPA}.}. 

The scalar field, $\Phi$, is a $2\times2$ matrix with $\Phi=\big{(} \tilde \phi \,| \,\phi\big{)} \, ,\tilde \phi=i\tau^2\phi^\star$ and $\phi$ an iso-doublet of complex scalar fields, that feels U$_Y(1)$ and SU(2)$_L$, but neither SU($N_c=3$) nor  SU($N_T=3$), gauge interactions. Despite the appearances, $\Phi$ is not the Higgs field, but rather an effective way to describe an UV completion of the model symmetric under the $\chi_L\times\chi_R$ global transformations defined below in eq.~(\ref{CHILR}).

For the SU(2)$_L$ SM matter doublets we use the notation $q_L=(u_L,d_L)^T$ and $\ell_L=(\nu_L,e_L)^T$, while {\it Right}-handed components ($q^u_R$, $q^d_R$ and $\ell^u_R\equiv \nu_R$, $\ell^d_R\equiv e_R$) are SU(2)$_L$ singlets. However, since for the moment we do not remove the {\it up}-{\it down} weak isospin degeneracy, in the fundamental Lagrangian we have used a ``doublet-like'' notation also for the {\it Right}-components of standard matter fermions. A similar notation is used for $Q$ and $L$ Tera-fermions. 

It is interesting to note that the form of the $d=6$ Wilson-like terms is completely fixed by symmetry and dimensional arguments plus the requirement of invariance under the independent constant shifts of fermionic fields $f(x)\to f(x) + {\mbox{const}}\, , \bar f(x)\to \bar f(x) + \overline{\mbox{const}}\, , f=q,\ell,Q,L$, in the limit $g_{Y,w,s,T} \to 0$~\cite{Golterman:1989df}. It follows that with the SM hypercharge assignments, where $y_{\nu_R}=0$ (see Table~\ref{tab:standard} in Appendix~\ref{sec:APPA}), the terms in eqs.~(\ref{LK}) and~(\ref{LW}) involving the lepton $\ell_R$ take the form
\beqn
\hspace{-.8cm}&&\bullet \,\ell_R \,\, /\!\!\!\!  {\cal D}^{B} \ell_R = \bar \ell_R^{u} \,\, /\!\!\! \partial  \,\ell_R^{u}+\bar \ell_R^{d} \,\, /\!\!\!\!  {\cal D}^{B} \ell_R^{d}= \bar \nu_R \,\, /\!\!\! \partial  \,\nu_R+\bar e_R \,\, /\!\!\!\!  {\cal D}^{B} e_R\label{SIMPK}\, ,\\
\hspace{-.8cm}&&\bullet \,\frac{{b^2}}{2}{\rho_\ell}\,\big{(}\bar \ell_L{\overleftarrow {\cal D}}\,^{BW}_\mu\Phi {\cal D}^{B}_\mu \ell_R+{\mbox{hc}}\big{)}= \nn\\
\hspace{-.8cm}&&\quad=\frac{{b^2}}{2}{\rho_\ell}\,\big{(}\bar \ell_L{\overleftarrow {\cal D}}\,^{BW}_\mu\widetilde\phi\, \partial_\mu \ell^u_R+{\mbox{hc}}\big{)} + \frac{{b^2}}{2}{\rho_\ell}\,\big{(}\bar \ell_L{\overleftarrow {\cal D}}\,^{BW}_\mu\phi {\cal D}^{B}_\mu \ell^d_R+{\mbox{hc}}\big{)} =\nn\\
\hspace{-.8cm}&&\quad=\frac{{b^2}}{2}{\rho_\ell}\,\big{(}\bar \ell_L{\overleftarrow {\cal D}}\,^{BW}_\mu\widetilde\phi\, \partial_\mu \nu_R+{\mbox{hc}}\big{)} + \frac{{b^2}}{2}{\rho_\ell}\,\big{(}\bar \ell_L{\overleftarrow {\cal D}}\,^{BW}_\mu\phi {\cal D}^{B}_\mu e_R+{\mbox{hc}}\big{)}  \, .\label{SIMPW}\
\eeqn
These equations show that $\nu_R=\ell^u_R$ is ``sterile'' and that the Wilson-like term associated with it is not able to give the neutrino a mass. So in the present formulation of the model neutrinos are exactly massless~\footnote{Although we do not discuss here the crucial question of how to give a non-vanishing mass to neutrinos, we note that in the scenario we are advocating in this investigation there exists a natural seesaw-like scale for neutrino masses, namely, $\Lambda_T^2/\Lambda_{GUT}\sim [(1\div10)\,{\mbox{TeV}}]^2/10^{17\div 18}\,{\mbox{TeV}}\sim 10^{-5}\div10^{-3}{\mbox{eV}}$, with $\Lambda_T\sim (1\div10)\,{\mbox{TeV}}$ and $\Lambda_{GUT}\sim 10^{17\div 18}\,{\mbox{TeV}}$ being reasonable estimates of the RGI scale of the theory (see sect.~\ref{sec:SPCS}) and unification scale of ref.~\cite{Frezzotti:2016bes}, respectively.}.

We end with the following ``amusing'' remark. If we take for the Tera-particles the hypercharge assignment of Table~\ref{tab:alternative} that gives unification of gauge couplings~\cite{Frezzotti:2016bes}, one finds that {\it Left}-handed components do not couple to $B$ but only to $W$, while the opposite is true for the {\it Right}-handed components, simply because {\it Left}-handed Tera-hypercharges are all zero.

\subsection{Symmetries of the Lagrangian}
\label{sec:SYLA}

Among other obvious symmetries, the Lagrangian~(\ref{LTOT}) is invariant under the (global) transformations $\chi_L\times \chi_R$, involving all fermions, $f=q,\ell,Q,L$, the $W$ bosons, and the scalar, $\Phi$, given by the equations  ($\Omega_{L/R} \in {\mbox{SU}}(2)$) 
\begin{eqnarray}
\hspace{-.5cm}&&\chi_L\times \chi_R =  [\tilde\chi_L\times (\Phi\to\Omega_L\Phi)]\times [\tilde\chi_R\times (\Phi\to\Phi\Omega_R^\dagger)] \label{CHILR}\\
\hspace{-2.cm}&&\qquad{\tilde\chi_{L}}: 
\left \{\begin{array}{l}     
f_{L}\rightarrow\Omega_{L} f_{L} \qquad \qquad\bar f_{L}\rightarrow \bar f_{L}\Omega_{L}^\dagger\, ,
\\
\,W_\mu\rightarrow \Omega_{L}W_\mu\Omega_{L}^\dagger\\ 
\end{array}\right . \label{GTWTL}\\
\hspace{-.5cm}&&\qquad\tilde\chi_{R} : \quad \,f_{R}\rightarrow\Omega_{R} f_{R} \, ,\qquad \quad\bar f_{R}\rightarrow \bar f_{R}\Omega_{R}^\dagger \, .\label{GTWTR}
\end{eqnarray}
As explained in refs.~\cite{Frezzotti:2014wja} and~(I), for generic values of the Yukawa couplings and $k_b$, the Lagrangian~(\ref{LTOT}) is not invariant under the ``chiral'' transformations $\tilde\chi_{L} \times \tilde\chi_{R}$, defined by eqs.~(\ref{GTWTL}) and~(\ref{GTWTR}), because of the presence of the ${\cal L}_{Wil}$, ${\cal L}_{Yuk}$ operators and the scalar kinetic term. However, invariance under $\tilde\chi_{L} \times \tilde\chi_{R}$ can be recovered (up to O($b^{2}$) terms) by enforcing the conservation of the $\tilde\chi_{L} \times \tilde\chi_{R}$ currents in the Wigner phase of the theory (where the scalar potential has a unique minimum). According to the general  strategy suggested in refs.~\cite{Bochicchio:1985xa,Testa:1998ez} and adapted to the present situation in refs.~\cite{Frezzotti:2014wja} and~(I), this occurs at the critical values, $\eta_{f\,cr}$ and $k_{b\,cr}$, of the parameters $\eta_f$ and $k_b$ that solve the equations ($f=q,\ell,Q,L $)
\beqn
&&\eta_f=\bar \eta_{f}^L\Big{(} \{g\};\{\eta\},\{\rho\};k_b\Big{)}\, , 
\label{CRETAU}\\
&&\eta_f=\bar \eta_{f}^R\Big{(} \{g\};\{\eta\},\{\rho\};k_b\Big{)} \, ,
\label{CRETAD}\\
&&k_b=\bar k^{L}_b\Big{(} \{g\};\{\eta\},\{\rho\};k_b\Big{)} \, ,
\label{CRKBL}
\eeqn
where $\{g=g_Y,g_w,g_s,g_T\}$ is the set of the gauge couplings of the theory. The superscript $L$ ($R$) refer to the tuning condition derived by enforcing the conservation of the $\tilde\chi_L$ ($\tilde\chi_R$) current, see eq.~(\ref{GTWTL}) (eq.~(\ref{GTWTR})). With $\{\eta\}$ and $\{\rho\}$ we denote the set of Yukawa and $\rho$ parameters associated to the fermions $f=q,\ell,Q,L$. The functions $\bar \eta_{f}^{L}$ and $\bar \eta_{f}^{R}$ in the r.h.s.\ of the eqs.~(\ref{CRETAU}) and~(\ref{CRETAD}) are the mixing coefficients of the ($\tilde\chi_L\times\tilde\chi_R$ rotations of the) $d=6$ Wilson-like operator of the $f$ fermion with the ($\tilde\chi_L\times\tilde\chi_R$ variation of the) corresponding $d=4$ Yukawa term, while $\bar k^L_b$ is the mixing coefficient with the ($\tilde\chi_L$ rotation of the) scalar kinetic operator.

Details of the procedure which leads to the tuning conditions~(\ref{CRETAU})--(\ref{CRKBL}) are given in Appendix~A of~(I), where we also prove that, owing to the exact $\chi_L\times\chi_R$ invariance, solving the set of eqs.~(\ref{CRETAU}) and~(\ref{CRKBL}) also solves the eqs.~(\ref{CRETAD}), despite the lack of parity invariance of the fundamental Lagrangian~(\ref{LTOT}).

\section{NP elementary particles mass generation}
\label{sec:MASS}

Generalizing the diagrammatic analysis developed in~\cite{Frezzotti:2014wja} and further expanded in~(I), one can extend the derivation of the parametric expression of quark, Tera-quark and $W$ masses given in~(I), to leptons, Tera-leptons and $B$. 

To this end, first of all, we must determine the list of the formally O($b^2$) $d=6$ NP Symanzik operators that need to be taken into account to describe the NP effects coming from the spontaneous breaking of the recovered $\tilde\chi_L\times\tilde\chi_R$ (chiral) symmetry. Naturally, the list given in eqs.~(3.11)--(3.13) of~(I) must be extended to include also $d=6$ operators involving leptons, Tera-leptons and hypercharge bosons. The relevant operators are reported below
\beqn
\hspace{-1.4cm}&&\gamma_{\bar Q Q}^T O_{6,\bar Q Q}^T =  r_{\bar Q Q}^T  b^2\Lambda_T \,\alpha_T |\Phi| \,\Big{[} \bar Q_L \,\, /\!\!\!\!  {\cal D}^{AGBW} Q_L+\bar Q_R \,\, /\!\!\!\!  {\cal D}^{BAG} Q_R \Big{]} \label{OTT} \\
\hspace{-1.4cm}&&\gamma_{\bar LL}^T O_{6,\bar L L}^T =r_{\bar LL}^T b^2 \Lambda_T\,\alpha_T|\Phi| \,\Big{[}\bar L_L\,\, /\!\!\!\!  {\cal D}^{GBW} L_L+\bar L_R \,\, /\!\!\!\!  {\cal D}^{GB} L_R \Big{]} \label{OLL} \\
\hspace{-1.4cm}&&\gamma_{AA}O_{6,AA} =r_{AA} b^2\Lambda_T\,g_s^2|\Phi|\, F^A\!\cdot\!F^A\label{OAA}\\
\hspace{-1.4cm}&&\gamma_{GG}^Q O_{6,GG}^Q =r_{GG}^Q b^2\Lambda_T\,g_T^2 |\Phi| \,F^G\!\cdot\!F^G \label{OGGQ} \\
\hspace{-1.4cm}&&\gamma_{GG}^L O_{6,GG}^L = r_{GG}^L  b^2 \Lambda_Tg^2_T\, |\Phi|\, F^G\!\cdot\!F^G \label{OGGL} \\
\hspace{-1.4cm}&&\gamma_{BB}^Q O_{6,BB}^Q\, =r_{BB}^Q  b^2 \Lambda_T\, g^2_Y|\Phi| \,F^B\!\cdot\!F^B \label{OBBQ}\\
\hspace{-1.4cm}&&\gamma_{BB}^L O_{6,BB}^L\, = r_{BB}^L b^2 \Lambda_T\, g^2_Y|\Phi| \,F^B\!\cdot\!F^B\, , \label{OBBL}
\eeqn
where the coefficients $r$'s are in principle computable, numerical constants, functions of the $\rho_f$ parameters, $N_c$ and $N_T$, with $r_{BB}^Q$ and $r_{BB}^L$ proportional to the square of the $Q$ and $L$ hypercharges, respectively. The expression of the various covariant derivatives can be obtained from eq.~(\ref{DER}). 

In Appendix~\ref{sec:APPB} we discuss the physical origin of the NP Symanzik operators~(\ref{OTT})--(\ref{OBBL}), displayed in figs.~\ref{fig:fig12} and~\ref{fig:fig13}. We argue that the ``optimal'' choice to parametrize the gauge coupling dependence (i.e.\ the choice that correctly  takes into account the $1/4\pi$ factors arising in loop integrals) is to have $\alpha$ in eqs.~(\ref{OTT}) and~(\ref{OLL}) and $g^2$ in the rest of the operators in the list. For phenomenology this choice is important (see sect.~\ref{sec:SPCS}) as there is a non-irrelevant numerical difference between the two alternatives.

As discussed in~(I), we want to stress also here that NP operators like those in the equations from~(\ref{OTT}) to~(\ref{OBBL}), though formally of O($b^2$), cannot be neglected if we want to get the correct description of the physics of the model, including features of NP origin. The reason is that, as we have seen in refs.~\cite{Frezzotti:2014wja} and~(I), when these operators are inserted in the correlators of the fundamental Lagrangian~(\ref{LTOT}), the IR $b^2$-factors they are proportional to turn out to be exactly compensated by the UV power divergencies of the loop integrals, finally yielding a non-vanishing, finite result. 

According to refs.~\cite{Symanzik:1983gh,Symanzik:1983ghh}, the proper way to keep all these NP effects correctly into account is to construct an ``augmented Lagrangian'' by adding to the fundamental Lagrangian the linear combination of the NP operators~(\ref{OTT})--(\ref{OBBL}). This approach was first employed in the simpler cases described in ref.~\cite{Frezzotti:2014wja} (see footnote~12) and extended in~(I) (see eq.~(3.14) there).

\begin{figure}[htbp]
\center\includegraphics[scale=0.5,angle=0]{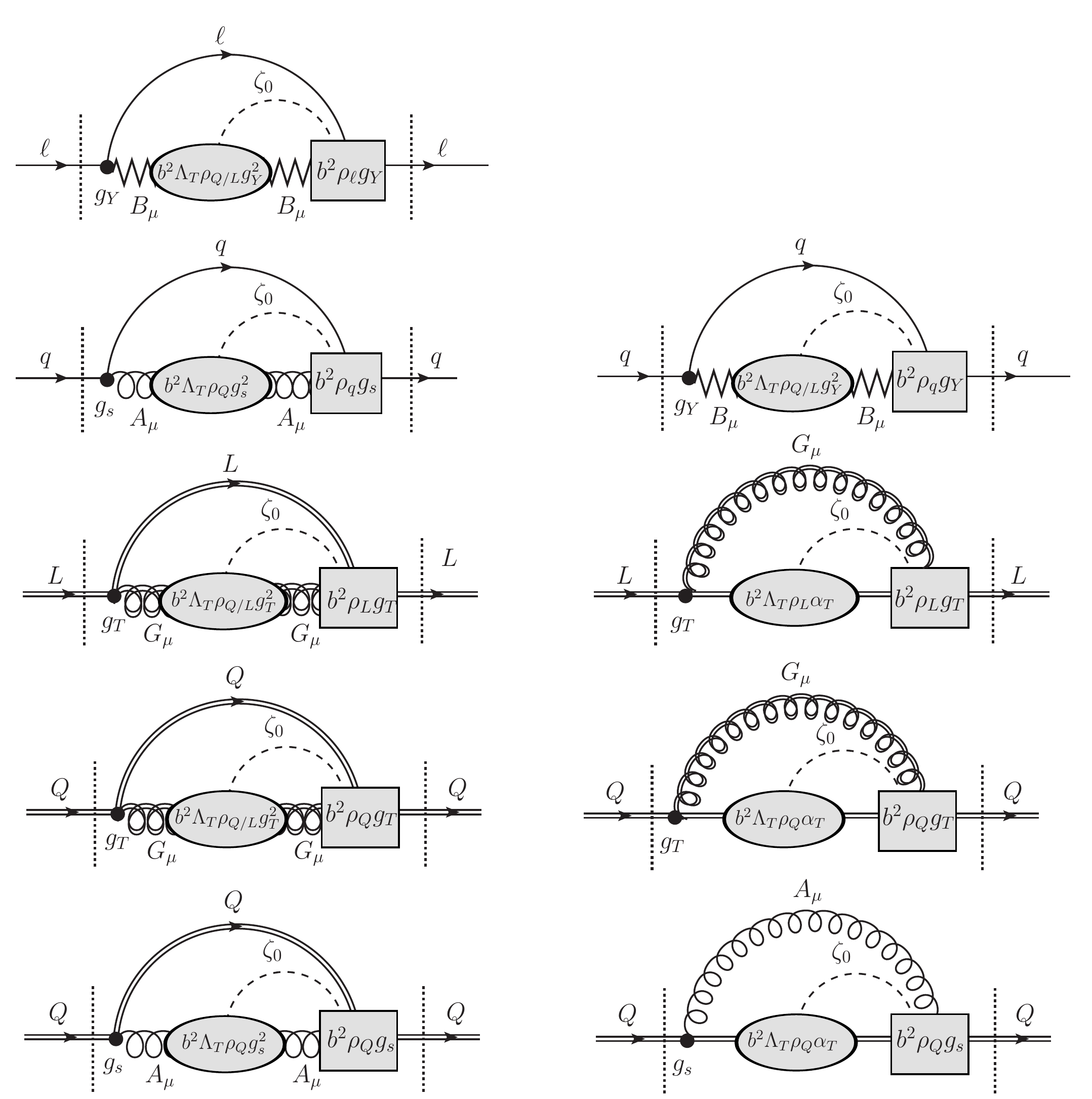}
\caption{\small{Examples of lowest loop order NP self-energy diagrams contributing (from top to bottom) to the masses of leptons, quarks, Tera-leptons and Tera-quarks, respectively. The blobs represent  the appropriate Symanzik operators among those displayed in equations from~(\ref{OTT}) to~(\ref{OBBL}), and the boxes the insertion of the Wilson-like vertices necessary to close the loops. The notation $Q/L$ means that either Tera-quarks or Tera-leptons run inside the blob. Single lines represent SM particles, double lines Tera-particles.}}
\label{fig:fig1}
\end{figure}

\begin{figure}[htbp]
\center\includegraphics[scale=0.5,angle=0]{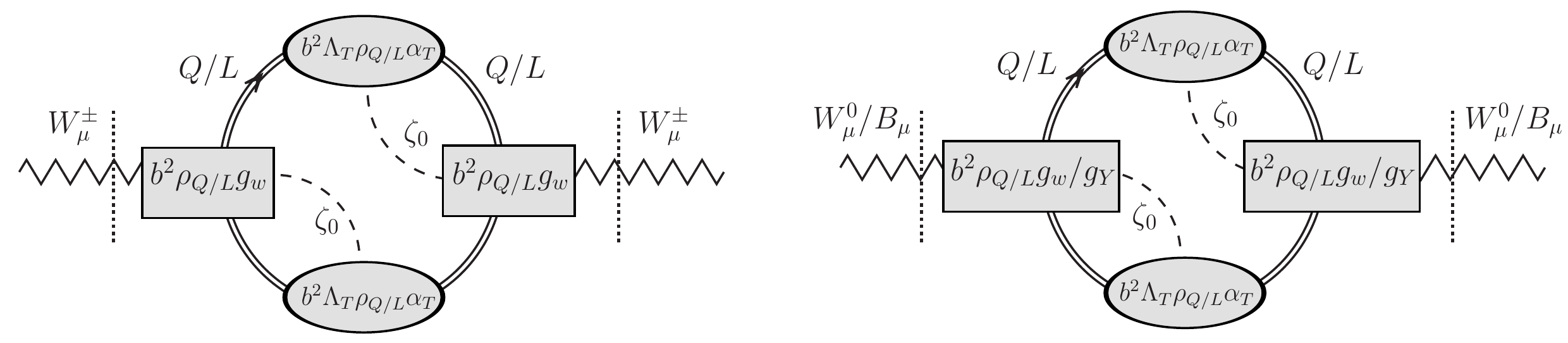}
\caption{\small{Examples of lowest loop order NP self-energy diagrams contributing to $W$ and $B$ masses. The notation $W^0_\mu/B_\mu$ means that either a neutral $W$ or a U(1)$_Y$} boson can be emitted or absorbed, accompanied by a $g_w$ or $g_Y$ gauge coupling, respectively. The rest of he notation is as in fig.~\ref{fig:fig1}}
\label{fig:fig11}
\end{figure}

Naturally the augmented Lagrangian gives rise to a completely new class of contributions.\ In particular, closing the external legs loop with appropriate Wilson-like terms, one can generate self-energy diagrams of NP nature.\ At the lowest loop order one finds the typical (amputated) self-energy diagrams displayed in the pictures of figs.~\ref{fig:fig1} and~\ref{fig:fig11}~\footnote{The vertical dotted lines mean amputation of the external leg. We have nevertheless left explicit ingoing and outgoing lines to help the reader recognizing the particle one is referring to.}. In fig.~\ref{fig:fig1} from top to bottom we show the diagrams that provide NP masses to leptons, quarks, Tera-leptons, Tera-quarks, respectively. The diagrams that give mass to charged $W$'s and neutral EW bosons are displayed in fig.~\ref{fig:fig11} where the right-most figure is supposed to compactly represents the four amputated propagators $W^0W^0$, $W^0B$, $BW^0$ and $BB$.

In the figs.~\ref{fig:fig1} and~\ref{fig:fig11} the blobs with external Tera-fermion--anti-Tera-fermion field (and scalar) legs stand for the operators depicted in fig.~\ref{fig:fig12}, while the blobs with external gauge field (and scalar) legs stand for the operators depicted in fig.~\ref{fig:fig13}.


Two observations are in order here. First of all, as we said, all these diagrams are finite, because of the exact compensation between the UV divergency of the multi-loop integrals and the $b^2$ factors brought in by the insertion of the~(\ref{OTT})--(\ref{OBBL}) operators and Wilson-like terms. This compensation is essentially based on dimensional arguments, thus we expect it to hold at all loops. Secondly, the fact that the masses come from the region of phase-space where all momenta are O($b^{-1}$) is telling us that what the self-energy diagrams represent are the masses at asymptotically large values of the cutoff, or better, if the gauge couplings unify (like in the case discussed in ref.~\cite{Frezzotti:2016bes} that we want to consider here), at the GUT scale. This remark will be crucial in the phenomenological considerations presented in sect.~\ref{sec:SPCS}.

An analysis of the expressions of the building blocks making up the diagrams in figs.~\ref{fig:fig1} and~\ref{fig:fig11} yields the following parametric mass formulae where we only expose the dependence upon $\Lambda_T$ and the gauge couplings~\footnote{The multiple gauge coupling dependence displayed in $C_q$, $C_L$ and $C_Q$ is in correspondence with the multiple diagrams contributing to $m_q$, $m_L$ and $m_Q$ reported in fig.~\ref{fig:fig1}.}
\beqn
\begin{array}{lllll}
&\hspace{-.4cm}{m_\ell}= C_\ell\,\Lambda_T\, , &C_\ell=c_\ell\,{\mbox{O}}\,({\alpha_Y\, g_Y^2}) & & \\
&\hspace{-.4cm}{m_q}= C_q\,\Lambda_T\, , &C_q=c_q\,{\mbox{O}}({\alpha_s\, g_s^2,\alpha_Y\, g_Y^2}) &&\\
&\hspace{-.4cm}{m_L} = C_L\,\Lambda_T\, , &C_L=c_L\,{\mbox{O}}({\alpha_T\,g_T^2, \alpha_T^2}) && \\
&\hspace{-.4cm}{m_Q} = C_Q\,\Lambda_T\, , &C_Q=c_Q\,{\mbox{O}}({\alpha_T\,g_T^2,\alpha_T^2,\alpha_s\, g_s^2,\alpha_T\,\alpha_s} ) & &\\
&&&&\\
&\hspace{-.4cm}{M_{W^\pm}} \!= C_{W^\pm}\,\Lambda_T\, ,\qquad&C_{W^\pm}={g_w} c_w \, ,\qquad\quad \,\,c_w=k_w{\mbox{O}}({\alpha_T}) &&\\
&\hspace{-.4cm}{M_{Z}} = C_{Z^0}\,\Lambda_T\, ,\qquad&C_{Z^0}=\sqrt{g^2_w+g^2_Y}\,c_w 
&&\\
&\hspace{-.4cm}{M_{A^0}} = 0\, . &&\\
\end{array}
\label{MASSES} 
\eeqn
All masses are proportional to $\Lambda_T$ times powers of the gauge couplings. In the case of the fermion masses the factors $g^2$ come from the gauge coupling dependence of the Symanzik operator from~(\ref{OGGQ}) to~(\ref{OBBL}), while the factors $\alpha$ come either from the operators~(\ref{OTT})-(\ref{OLL}) or from closing the loop with a gauge boson extracted from a Wilson-like vertex.\ In the case of $M_W$ the factors $\alpha_T$ come from the operators~(\ref{OTT}) and~(\ref{OLL}), while the explicit $g_w,g_Y$ factors represent the coupling of the external EW boson legs to the Tera-fermions. With these specific choices we (believe we) have correctly taken care of the $1/4\pi$ factors arising in loop integrals. 

Notice that, owing to custodial symmetry, which is unbroken at the leading order in the EW interactions, the diagonalization of the self-energy matrix represented by the right diagrams of fig.~\ref{fig:fig11} yields, exactly like in the SM, a massive $Z$ boson and a massless photon. 

A further interesting observation (see subsect.~\ref{sec:MTTF} of Appendix~\ref{sec:APPC}) is that in the limit of large $N_c$ and $N_T$, all the $C$ coefficients in eq.~(\ref{MASSES}) are non-vanishing and finite. In the same limit, as we discussed at the end of Appendix~\ref{sec:APPC} (see eqs.~(\ref{MQUARK})--(\ref{MWW})), one also recognizes that $c_Q$ and $c_W$ are $\rho$-independent quantities.

The gauge coupling dependence exposed in eqs.~(\ref{MASSES}) is what results if we assume that the Wilson-like terms of all elementary particles are $d=6$ operators. For the need of phenomenology this may not be the best  choice. As illustrated in sect.~\ref{sec:SPCS}, we will stick to $d=6$ for the Wilson-like operators of Tera-particles that are responsible for the generation of the fundamental NP Symanzik operators~(\ref{OTT})--(\ref{OBBL}), but we will leave us the freedom of appropriately choosing the dimension of the Wilson-terms of SM fermions to best match the phenomenological values of their masses.  This freedom does not affect the expression of the EW boson masses because, as the diagrams in fig.~\ref{fig:fig11} show, only the Wilson-like operators of Tera-particles matter.

\section{Universality and predictive power}
\label{sec:RDU}

\subsection{The $\rho$ dependence}
\label{sec:RDEP}

In none of the formulae of the previous section (eqs.~(\ref{OTT})--\ref{OBBL}) and eq.~(\ref{MASSES})) we have indicated the dependence on the $\rho$ parameters. For completeness we briefly recall here the argument we presented in sect.~5.1 of~(I) to argue that the physics of the model has a mild dependence (if any at all) upon the $\rho$'s.

We first observe that in the diagrams of  figs.~\ref{fig:fig1} and~\ref{fig:fig11} the $\rho$ dependence arises from the product of pairs of $\rho$ factors coming from the NP Symanzik operators and Wilson-like vertices, times one power of $k_{b\,cr}^{-1}$ from the $\zeta_0$ propagator joining the two vertices. Since $k_{b\,cr}$ is (up to radiative corrections) a quadratic form in the $\rho$'s (see eq.~(\ref{KBCR}) of Appendix~\ref{sec:APPC}) and we have two factors of $\rho$ for each $\zeta_0$ propagator, we immediately see that all the NP self-energy diagrams are only functions of $\rho$ ratios. Actually, one can show (see Appendix~\ref{sec:APPC} below) that this property is exact and holds for all physical observables. 

We have already observed in the second paragraph of sect.~5.1 of~(I) that the dependence of physical quantities on $\rho$ ratios could be mitigated or even eliminated by conjecturing that some symmetry exists which, putting constraints on the $\rho$'s, restricts their ratio variability range. For instance, at least in the situation where all Wilson-like terms are $d=6$ operators (like in the case of the Lagrangian~(\ref{LTOT})), an extreme but appealing situation would the one in which all the $\rho$'s are equal because of some GUT symmetry. In this case all the $\rho$ ratios would be equal to unit.

\subsection{Universality}
\label{sec:UNIV}

We already observed in sect.~5 of~(I) that the issue of the $\rho$ dependence of the NP-ly generated masses impacts on the problem of universality because it looks that masses depend on the values we assign to their ratios. Actually, as explained there, the situation is somewhat more complicated than that. One can see that masses also depend on the detailed form of the ``irrelevant'' (chiral breaking) Wilson-like terms. In particular, generically the larger is the dimension of the Wilson-like operators the higher will be the power of the gauge couplings controlling the leading behaviour of the coefficients $C_f \,(f=q,\ell,Q,L)$ and $c_w$.

However (see sect.~5.1 of~(I)) rather than a problem for universality, this last fact might provide a handle to understand the family mass ranking, from heavier to lighter, as associated to Wilson-terms of different dimensions, from smaller to larger. At the moment we have little understanding of how it could be possible to rigorously implement this intriguing idea in the present formalism. However, in sect.~\ref{sec:SPCS} we show that, ignoring theoretical subtleties, we get very interesting mass ``postdictions'', if we give us the liberty of appropriately choosing the dimension of certain Wilson-like operators, like the ones associated to leptons. The same line of reasoning can be usefully extended to deal with the problem of understanding the mass splitting of the top--bottom weak iso-doublet. The whole issue is under intensive study from our side. 

\subsection{Some wild speculations}
\label{sec:ASPE}

Before continuing, we would like to pause for a moment and offer to the reader a couple of somewhat speculative observations concerning the structure of the bSMm we are developing and the role of EW interactions for the consistency of the model. We would like to argue that the existence of EW interactions is essential for the viability of the construction as an elementary particle theory. In fact, on the one hand, weak interactions allow getting rid of the radial component of the primitive scalar field $\Phi$ (as shown in Appendix~\ref{sec:APPB} of~(I)), thus providing a way to circumvent the Higgs mass tuning problem. On the other, in this approach hypercharge interactions are necessary to give mass to leptons, as weak interactions alone cannot do the job.

\section{Some phenomenological considerations}
\label{sec:SPCS}

The key questions that we need to address in order to put the present model on a solid basis and make it useful for phenomenology are 1) how we can make contact between the theoretical mass estimates provided by the formulae~(\ref{MASSES}) and the phenomenological values of elementary particle masses and 2) whether we can access the value of the scale of the new interaction from the knowledge of the available low energy physics. It goes without saying that both are very difficult problems of which we have only a very limited understanding. 

\subsection{Running of fermion masses}
\label{sec:RFM}

To deal with the question of fermion masses, a crucial question is to understand which scale the masses extracted from the evaluation of the NP self-energy diagrams of fig.~\ref{fig:fig1} refer to. As we have argued in~(I), the answer follows from the observation, we have already made a number of times, that the NP self-energy diagrams (like those in figs.~\ref{fig:fig1}) and~\ref{fig:fig11}) are finite and non-vanishing because of the exact compensation between IR $b^2$-factors and the UV power divergent behaviour of the loop integrals. This means that the region of momenta responsible for the emergence of masses is where all loop momenta are O($b^{-1}$). As a result, the NP running masses produced by these diagrams should be seen as evaluated at asymptotically large UV cut-off, or more realistically, if the theory unifies, at the unification scale, $\Lambda_{GUT}$. In the next subsections, exploiting this observation, we will work out from eqs.~(\ref{MASSES}) some crude phenomenological estimates of $\Lambda_T$ and elementary particle mass ratios. 

\subsection{Mass estimate strategy}
\label{sec:RFT}

In order to make contact between the mass estimates provided by the eqs.~(\ref{MASSES}) and the actual values of quark and $W$ masses, we remark that, if the mass generation mechanism described in this paper is realized in a model with a particle content that yields gauge coupling unification at some very high scale, say $\Lambda_{GUT}$ (like the model discussed in ref.~\cite{Frezzotti:2016bes}), then at that scale all fermion masses will be close to each other, owing to the fact that in the NP mass formulae~(\ref{MASSES}) the gauge couplings are all evaluated at the scale $\Lambda_{GUT}$ where they are essentially all equal. The evolution of running masses from $\Lambda_{GUT}$ down to the energy scale of the order of $\Lambda_T$ (or lower) will be different for different fermion species, reflecting the RGI evolution of the fermion bilinear operator to which each mass is coupled and the different running of the coupling of the gauge interactions each fermion is subjected to. One can hope to get in this way the correct phenomenological order of magnitude of elementary particle masses. 

The well-known formula that describes the fermion mass running from $\Lambda_1=\Lambda_{GUT}$ down to a scale $\Lambda_2$, that we take somewhat larger than $\Lambda_{T}$ (see below), and equal, as a try, to 5~TeV, at leading order reads  
\begin{equation}
 {m}_f(5~{\mbox{TeV}})  =  {m}_f(\Lambda_{GUT})
\prod_{p=Y,w,s,T} \Big[\frac{\alpha_p(5~{\mbox{TeV}})}{\alpha_p(\Lambda_{GUT}) }\Big]^{ \gamma_{0p}^{f} / 2\beta_{0p} } \, , \label{OLRF}
\end{equation}
where the product is over all the interactions felt by the fermion $f$.  As we said, we shall consider the framework of the fully unified gauge coupling model  worked out  in ref.~\cite{Frezzotti:2016bes}, where the unifying couplings $g_n,n=1,2,3,4$ are related to the physical couplings by the relations~\cite{Frezzotti:2016bes} 
\begin{equation}
g_1^2= \frac{4}{3} g_Y^2 \,\quad g_2^2=g_w^2\,,\quad g_3^2=g_s^2\,,\quad g_4^2=  \frac{8+N_S}{12} g_T^2 \, .
\label{UNIF2}
\end{equation}
For the first coefficient of the $\beta$ functions one finds~\cite{Frezzotti:2016bes} 
\beqn
\hspace{-1.2cm}&&\beta_{0T}=\frac{11}{3}N_T-\frac{4}{3}(N_c+1) \, ,\label{OLBSMM1}\\
\hspace{-1.2cm}&&\beta_{0s}=\frac{11}{3}N_c-\frac{4}{3}(N_T+N_f)\, , \label{OLBSMM2}\\
\hspace{-1.2cm}&& \beta_{0w}=2 \frac{11}{3}-\frac{1}{3}N_f(N_c+1)-\frac{1}{3} N_T(N_c+1) \, , \label{OLBSMM3}\\
\hspace{-1.2cm}&& \beta_{0Y}=-\frac{2}{3}\left[\left(\frac{22}{36}N_c+\frac{3}{2} \right)N_f+\frac{1}{2} N_T(N_c+1)\right]\, .\label{OLBSMM4}
\eeqn
Unification of U$_Y$(1), SU$_L$(2) and SU($N_c$) occurs with the (essentially unique) choice $N_T=N_c=N_f=3$. To get unification with also Tera-strong interactions the best choice is to take $N_S=5$. One should notice that in the present unifying scheme $\beta_{0w}$ is negative so, unlike the case of the SM, weak interactions are not asymptotically free. From~(\ref{OLBSMM3}) one finds $\beta_{0w}=-2/3$ leading to a very slow $g_w^2$ running.
 
The running of masses under EW interactions can be estimated to gives only some $10\div 20$\,\% contribution in eq.~(\ref{OLRF}). We will then neglect them in our calculations below, as they are well below the accuracy of our approximations. Focussing on strong and Tera-strong running effects, we get from eqs.~(\ref{OLBSMM1})--(\ref{OLBSMM2}) 
\beqn
&&\beta_{0s}= 3 \, ,\label{B5}\\  
&&\beta_{0T}= \frac{17}{3}  \, , \label{B6}
\eeqn
For the 1-loop  mass anomalous dimensions one has~\cite{PDG}
\beqn
\hspace{-1.cm}&&\gamma_{0s}^{Q}= \gamma_{0s}^{L}=\gamma_{0s}^{q}= 8\, ,\label{G5}\\ 
\hspace{-1.cm}&&\gamma_{0T}^{Q}= \gamma_{0T}^{L}= 8\, . \label{G6}
\eeqn
We determine the values of the gauge couplings at the scale of 5~TeV in the mass independent $\overline {\mbox{MS}}$ scheme from the plot reported in fig.~6 of ref.~\cite{Frezzotti:2016bes}. We find $\alpha_s(5~{\mbox{TeV}}) \sim {1/13}$ and $\alpha_T(5~{\mbox{TeV}}) \sim 2$. Taking $N_S=5$, we obtain from  eq.~(\ref{UNIF2})
\beqn
&&\alpha_3(5~{\mbox{TeV}})=\alpha_s(5~{\mbox{TeV}}) \sim\frac{1}{13}\, ,\label{UNIF5}\\ 
&&\alpha_4(5~{\mbox{TeV}})=\frac{8+N_S}{12} \alpha_T(5~{\mbox{TeV}}) = \frac{13}{12}\, \alpha_T(5~{\mbox{TeV}}) \sim  \frac{13}{12} \cdot 2\, .
\label{UNIF6}
\eeqn
On the basis of these observations, we propose to estimate the value of Tera-fermion masses and the masses of the fermion of the heaviest SM family in the $\overline {\mbox{MS}}$ scheme, starting from the asymptotic formulae
\beqn
&&m_f(\Lambda_{GUT})=C_f\, \overline \alpha_s^{\,u_f}\, \overline g^2_s \Lambda_T\, , \,\,\quad f=t,b\, ,\label{ASFSMF}\\
&&m_\tau(\Lambda_{GUT})=C_\tau\,\overline \alpha_Y^{\,u_\ell}\, \overline g^2_Y\Lambda_T\, ,\label{ASFSMLL}\\
&&m_Q(\Lambda_{GUT})=C_Q\,\overline \alpha_T^{\,u_Q}\, \overline g_T^2 \Lambda_T\, ,\label{ASFSMT} \\
&&m_L(\Lambda_{GUT})=C_L\,\overline \alpha_T^{\,u_L}\, \overline g_T^2 \Lambda_T\, ,\label{ASFSML} 
\eeqn
where over-lining means that the gauge couplings are taken at the scale $\Lambda_{GUT}$, where $g_1^2,g_2^2,g_3^2,g_4^2$ (see eq.~(\ref{UNIF2})) unify~\footnote{Actually, looking at eqs.~(\ref{MASSES}), we see that the dependence of masses on the gauge couplings is a bit more complicated that what it is indicated in eqs.~(\ref{ASFSMF})--(\ref{ASFSML}). However, these modulation effects can be neglected as they are well within the level of approximations of the numerical estimates we present in sect.~\ref{sec:MEN}.}. We recall that the factors of the kind $\overline\alpha^{\,u}$ comes either from the insertion of the operators~(\ref{OTT}) and~(\ref{OLL}) and/or from the loop integration involving the Wilson-like term associated to the fermion, while the factors $\overline g^2$ come from the insertion of the operators~(\ref{OAA})--(\ref{OBBL}) (see also fig.~\ref{fig:fig13} in Appendix~\ref{sec:APPB}). 

For the numerical values of the physical gauge couplings $\overline \alpha$'s, at the GUT scale we obtain
\beqn
\hspace{-.4cm}&\overline \alpha_Y \sim\dfrac{3}{4}\cdot\dfrac{1}{28}=\dfrac{3}{112}\qquad \quad\overline \alpha_w\sim\dfrac{1}{28}\qquad\qquad\overline \alpha_s \sim \dfrac{1}{28} \qquad\quad \overline \alpha_T\sim \dfrac{12}{13}\cdot\dfrac{1}{28}=\dfrac{12}{364}\, .\label{COG1}
\eeqn
These numbers are extracted from the unifying behaviour displayed in fig.~6 of ref.~\cite{Frezzotti:2016bes} with $N_S=5$, taking into account the factors in eq.~(\ref{UNIF2}). For the corresponding gauge couplings we get
\beqn
\hspace{-1.2cm}&\overline g_Y \sim 0.58 \qquad
\quad\quad  \overline g_w \sim 0.67\qquad
\quad\quad  \overline g_s \sim 0.67\qquad
\quad \quad \overline g_T  \sim 0.64\, .\label{COG2}
\eeqn
Concerning the integers $u$, we make the ``ad hoc'' choice, $u_t=u_Q=u_L=1$  for top, Tera-quarks and Tera-leptons, respectively. We recall that, as mentioned in sect.~\ref{sec:UNIV} the choice $u_f=1$ means that the Wilson-like operator of the fermion $f$ has dimension 6. For the $\tau$ lepton we take instead $u_\tau=2$ which means that for the $\tau$ lepton we assume we have a $d=8$ Wilson-like operator. A similar tuning is necessary if one wants to reproduce the mass of the lowest components of the top-bottom weak iso-doublet. As we shall see in sect.~\ref{sec:MBR}, one needs to take for the bottom quark $u_b=2$ to describe the top--bottom mass splitting.

\subsection{Mass estimate numerics}
\label{sec:MEN}

In order to get estimates for $\Lambda_T$ and masses we need to fix a scale. Recalling what we said at the end of Appendix~\ref{sec:APPC}, we decide to express everything in terms of the $W$ mass because the latter turns out to be given by the most robust of our mass formulae. First of all $M_W$ only depends upon the Wilson-like operators of the Tera-sector that we keep to be $d=6$ operators. Secondly, as we have mentioned in subsect.~\ref{sec:MTTF} of Appendix~\ref{sec:APPC}, one can recognize that $M_W$ has a rather weak dependence on $N_c$, $N_T$ and $\rho$ ratios, which actually disappears in the limit of large $N_c$, $N_T$. 

\subsubsection{$M_W$ mass}
\label{sec:MWM}

Using the fifth relation in eq.~(\ref{MASSES}) where, we recall, the factor $\overline\alpha_T$ comes from the insertion of the operators~(\ref{OTT}) and~(\ref{OLL}) and the factor $\overline g_w$ from the external weak coupling to which the $W$ is attached, our theoretical estimate for the $W$ pole mass gives 
\beq
M_W=k_w \overline \alpha_T\, \overline g_w \Lambda_T
\label{MWT1}
\eeq
where, as we noticed, $k_w$ is an (almost) $N_c$, $N_T$ and $\rho$ independent numerical coefficient. Using the eqs.~(\ref{COG1})-(\ref{COG2}), from $M_W\sim 80$\,GeV, we can extract the value of the product $k_w \Lambda_T$ finding
\beq
80=k_w \overline \alpha_T\, \overline g_w \Lambda_T=k_w \frac{12}{364}\, 0.67\Lambda_T \quad \to\quad k_w \Lambda_T\sim 3.6~{\mbox{TeV}}\, .
\label{MWTL}
\eeq
Notice that, in fixing the $W$ mass, we have not included the effects of the loops that, just like in the SM, also here would give corrections to the $M_W$ tree level value~\cite{PESK}. Although these corrections are very small, they are crucial for the precision tests of the SM~\cite{Peskin:2017emn}. In the present context we ignore them as we are only interested in the order of magnitude of the bulk contribution to the $W$ mass, i.e.\ in the term that in our approach replaces the SM formula $M_W\sim g_w v$.

\subsubsection{Tera-fermion mass running}
\label{sec:MQR}

Using~(\ref{OLRF}) we obtain from the running of the Tera-quark mass
\beqn
&&m_Q(5~{\mbox{TeV}})=C_Q\overline \alpha_T\, \overline g^2_T \Lambda_T 
\Big{(}\frac{2}{12/364}\Big{)}^{8 \frac{3}{17}\frac{1}{2}} 
\Big{(}\frac{1/13}{1/28}\Big{)}^{\frac{8}{3}\frac{1}{2}} \sim \nn\\
&&\sim \frac{C_Q}{k_w} \frac{12}{364}\, (0.64)^2 \cdot 18.15 \cdot 2.78 \cdot 3600 \sim  \frac{C_Q}{k_w}\, 2500~{\mbox{GeV}}\, , \label{MQR31}
\eeqn
where the last two factors in the first line of eq.~(\ref{MQR31}) correspond to the running induced by Tera-strong and  strong interactions, respectively, and we have neglected EW running corrections.

From the running of the Tera-lepton mass we find (it is enough to drop the last factor in the first line of eq.~(\ref{MQR31}) which represents the contribution from the strong interaction running)
\beqn
&&m_L(5~{\mbox{TeV}})=C_L\overline \alpha_T\, \overline g^2_T \Lambda_T 
\Big{(}\frac{2}{12/364}\Big{)}^{8 \frac{3}{17}\frac{1}{2}} \sim  \frac{C_L}{k_w}\, 900~{\mbox{GeV}}\label{MLR1} \, .
\eeqn

\subsubsection{$m_t$ running}
\label{sec:MTR}

For the top mass we get
\beqn
 \hspace{-.8cm}&&m_t(5~{\mbox{TeV}})=C_t\,\overline \alpha_s\, \overline g_s^2 \Lambda_T 
\Big{(}\frac{1/13}{1/28}\Big{)}^{\frac{8}{3}\frac{1}{2}} \sim \frac{C_t}{k_w} \, \frac{1}{28}(0.67)^2 \cdot 2.78 \cdot 3600 \sim\frac{C_t}{k_w} \, 160~{\mbox{GeV}} \, ,\label{MQR1} 
\eeqn
where again we have neglected EW running factors.

\subsubsection{$m_b$ running}
\label{sec:MBR}

To describe the top-$b$ weak isospin breaking we conjecture that the Wilson-like operator associated to the $b$ quark is of dimension 8 giving, as we said below eq.~(\ref{COG2}), $u_b=2$. This leads in the $b$ mass formula to the presence of an extra $\overline \alpha$ factor with respect to the top mass formula, that we take to be $\alpha_Y(\Lambda_{GUT})= \overline \alpha_Y$, with the idea that the weak isospin splitting is related to EW interactions. Always neglecting EW running factors, we get 
\beqn
\hspace{-1.8cm}&&m_b(5~{\mbox{TeV}})=C_b\,\overline \alpha_s\overline g_s^2 \overline \alpha_Y\Lambda_T \Big{(}\frac{1/13}{1/28}\Big{)}^{\frac{8}{3}\frac{1}{2}}
\sim \frac{C_b}{k_w} \,\frac{1}{28} (0.67)^2 \frac{3}{112}\, \cdot 2.78 \cdot 3600\sim \frac{C_b}{k_w} \,4.3~{\mbox{GeV}} \, .
\eeqn

\subsubsection{$m_\tau$ mass}
\label{sec:MTAUR}

For the calculation of the $\tau$ lepton pole mass we will assume that the associated Wilson-like operator is of dimension 8 leading, as we said below eq.~(\ref{COG2}), to $u_\tau=2$. Thus also in the $\tau$ mass formula we have to include an extra $\overline \alpha$ factor with respect to the top mass formula, that we take to be $\alpha_Y(\Lambda_{GUT})= \overline \alpha_Y$, with the idea that leptons are lighter than quarks because they are only charged under EW interactions. Numerically from eq.~(\ref{ASFSMLL}) we obtain
\beqn
&&m_\tau \sim C_\tau \overline \alpha_Y^2\, \overline g_Y^2 \Lambda_T \sim \frac{C_\tau}{k_w} \left(\frac{3}{112}\right)^2(0.58)^2 \cdot 3600\sim \frac{C_\tau}{k_w} \,0.87~{\mbox{GeV}} \, .
\label{MTAUR2} 
\eeqn

\subsection{Summary \& comments}
\label{sec:COM}

In summary, with the admittedly {\it ad hoc} choices we have made and using the $W$ mass ($M_W\sim 80~{\mbox{GeV}}$) as an input, we have obtained the following numbers 
\beqn
&&k_w \Lambda_T \sim 3600~{\mbox{GeV}}\label{SUM1}\\
&& m_Q(5~{\mbox{TeV}}) \sim \frac{C_Q}{k_w} \,2500~{\mbox{GeV}}\label{SUM2}\\
&& m_L(5~{\mbox{TeV}}) \sim \frac{C_L}{k_w} \,900~{\mbox{GeV}}\label{SUM3}\\
&&m_t(5~{\mbox{TeV}}) \sim \frac{C_t}{k_w} \,160~{\mbox{GeV}}\\
\label{SUM6}
&& m_b(5~{\mbox{TeV}}) \sim \frac{C_b}{k_w} \, 4.3~{\mbox{GeV}}\label{SUM5}\\
&& m_\tau \sim \frac{C_\tau}{k_w} \,0.87~{\mbox{GeV}}\label{SUM4}
\eeqn
where the first equation provides an idea of the order of magnitude of $\Lambda_T$. Given the level of the approximations we had to inject in these estimates, where factors of 2 are out of control, if one imagines to take all constant ratios, $C_f/k_w$, equal to unit, the numbers above are even too good. Perhaps only the lepton masses seem to be a bit too low. Naturally, there wold be no problem in bringing the running fermion masses down to their respective self-consistent $\overline m_f$ mass scale.

\subsection{A (little) help from lattice simulations}
\label{sec:LF}

The key question is to what extent we can test the ability of our model in predicting elementary particle masses. Given the NP nature of the mechanism underlying the emergence of masses, the only practical way one can envision to perform checks of this type is to try to set up some kind of lattice Monte Carlo simulations. Naturally, simulating the whole theory defined by eq.~(\ref{LTOT}) appears to be an impossible task, because we would need to integrate over multiple sets of gauge fields and fermions. 

However, for the evaluation of some specific quantity an approximate lattice computational scheme can be set up. Luckily it turns out that this can be done for the especially important ratio $M_W/m_Q$, which we expect to find {\it dynamically} much smaller than unit, if the scheme we are proposing has to have any chance of being phenomenologically useful.  

Neglecting strong and EW virtual corrections, compared to Tera-strong effects, as well as the impact of Tera-leptons, one can imagine to extract $M_W$ by performing in the lattice regularized version of the toy-model introduced in ref.~\cite{Frezzotti:2014wja} (also discussed in sect.~2 of~(I) and already employed in the simulations described in ref.~\cite{Capitani:2019syo}) unquenched Monte Carlo simulations of the two point correlator $g_w^2\langle \widehat J_\mu^{i\,L}(p) \widehat J_\nu^{j\,L}(-p) \rangle$ where $\widehat J_\mu^{i\,L}(p)$ is Fourier transform of the conserved {\it Left}-handed current to which the external $W$ fields are coupled. In this framework quark and gluons should be interpreted as Tera-quarks and Tera-gluons. Naturally, extracting the value of the $W$ mass from this kind of simulations is expected to be numerically rather difficult.

In the same lattice framework one can compute the PCAC mass of the fermion, that we identify with $m_Q$. This calculation only accounts for the contribution to $m_Q$ coming from the diagrams in the fourth panel of fig.~\ref{fig:fig1}. Almost equal contributions can be estimated to come from the diagrams in the fifth panel, because at the GUT scale $\alpha_s$ and $\alpha_T$ are equal, apart from the small rescaling implied by eq.~(\ref{UNIF2}). 

\section{The QEL of the critical theory}
\label{sec:QEL}

The form of the QEL, $\Gamma_{cr}^{NG}$, that describes the physics of the critical theory in the NG phase, including the NP mass terms identified in sect.~\ref{sec:MASS}, is essentially dictated by geometrical considerations. In fact, it is highly constrained by dimensional and symmetry arguments, in particular by the invariance under the exact $\chi_L\times\chi_R$ symmetry and the observation that at the critical point neither the scalar field kinetic term nor the Yukawa terms, that both would break $\tilde\chi_L\times\tilde\chi_R$, should be present in $\Gamma_{cr}^{NG}$ (see~\cite{Frezzotti:2014wja} and~(I)). Focusing on its $d\leq 4$ part, the expression of the QEL of the theory is obtained by including all the operators of dimension $d\leq 4$, invariant under $\chi_L\times\chi_R$ that can be constructed in terms of the matter and gauge fields of the theory and the non-analytic field $U$, defined by the polar decomposition (meaningless in the Wigner phase)
\beq
\Phi = RU\, , \quad R=v+\zeta_0\, ,\quad U=\exp[i\vec\tau\,\vec \zeta/c_w\Lambda_T]\, ,\label{PHPH}
\eeq
where in standard notations, $v$ is the vev of the scalar. The choice of the (arbitrary) mass scale in the exponent has been conveniently made with an eye to eq.~(\ref{GW}), i.e.\ so as to have the NG bosons, $\zeta^i, i=1,2,3$, canonically normalized. On the basis of the constraints imposed by the above considerations, one obtains 
{\beqn
\hspace{-1.cm}&&{\Gamma}_{4\, cr}^{NG}(q,\ell,Q,L;\Phi;A,G,W,B)=
\frac{1}{4}\Big{(}F^A\!\cdot\! F^A\!+\! F^G\!\cdot\! F^G\!+\!F^W\!\cdot\! F^W\!+\!F^B\!\cdot\! F^B\Big{)}+\nn\\ 
\hspace{-1.cm}&&\qquad+\Big{[}\bar q_L \,\,/\!\!\!\! {\cal D}^{ABW} q_L+\bar q_R \,\,/\!\!\!\!  {\cal D}^{AB} q_R\Big{]}+
C_q\Lambda_T\,\Big{(} \bar q_L U q_R+\bar q_R U^\dagger q_L \Big{)}+\nn\\
\hspace{-1.cm}&&\qquad+\Big{[}\bar \ell_L \,\,/\!\!\!\!  {\cal D}^{BW} \ell_L+\bar \ell_R \,\,/\!\!\!\!  {\cal D}^{B} \ell_R\Big{]} +
C_\ell\Lambda_T\,\Big{(} \bar \ell_L U \ell_R+\bar \ell_R U^\dagger \ell_L \Big{)}+\nn\\
\hspace{-1.cm}&&\qquad +\Big{[}\bar Q_L\,\,/\!\!\!\! {\cal D}^{AGBW} Q_L+\bar Q_R\,\,/\!\!\!\! {\cal D}^{AGB} Q_R\Big{]}
+C_Q\Lambda_T\,\Big{(} \bar Q_L U Q_R+\bar Q_R U^\dagger Q_L \Big{)}+\nn\\
\hspace{-1.2cm}&&\qquad +\Big{[}\bar L_L\,\,/\!\!\!\! {\cal D}^{ABW} L_L+\bar L_R\,\,/\!\!\!\! {\cal D}^{AB} L_R\Big{]}
+C_L\Lambda_T\,\Big{(} \bar L_L U L_R+\bar L_R U^\dagger L_L \Big{)}+\nn\\
\hspace{-1.2cm}&&\qquad +\frac{1}{2}c_w^2\Lambda_T^2{\tr}\big{[}({\cal D}\,^{BW}_\mu U)^\dagger {\cal D}^{BW}_\mu U\big{]} \label{GW}\, .
\eeqn 
Expanding $U$ around the unit matrix we immediately have the mass identifications summarized in the formulae~(\ref{MASSES}). We want to stress again that in the present approach mass terms in ${\Gamma}_{cr}^{NG}$ have a somewhat unusual expression and a peculiar conceptual status. They appear as a sort of NP anomalies preventing the full recovery of the $\tilde\chi_L\times\tilde\chi_R$ symmetry.

We end by noting that actually there exist other operators invariant under $\chi_L\times\chi_R$. At $d=4$ we have the operator $\Lambda_s R\,{\tr}\big{[}({\cal D}\,^{BW}_\mu U)^\dagger {\cal D}^{BW}_\mu U\big{]}$ and the scalar potential ${\cal V}(\Phi)$. However, as we proved in the Appendix~B of~(I), in the presence of weak interactions, the restoration of $\tilde\chi_L\times\tilde\chi_R$ makes the first two operators to disappear from the QEL, while the (effective) singlet field $R=v+\zeta_0$ (eq.~(\ref{PHPH})) becomes infinitely massive and decouples (see also the remark in sect.~\ref{sec:ASPE} and Appendix~C in~(I)). This is the reason why there is no reference to the value of $v$ in ${\Gamma}_{cr}^{NG}$, hence in the physics described by the model in its NG phase.

\subsection{The effective low energy Lagrangian of the model and the SM}
\label{sec:QELSM}

Following the line of arguments given in~(I), for completeness we briefly recall here the main steps that allow establishing the not-unexpected (and welcome) similarity between the $d=4$ LEEL of our model valid for momenta$^2\!\ll\!\Lambda_T^2$ and the SM Lagrangian. 

In order to get the expression of the LEEL of the model in these kinematical conditions, one needs first of all to integrate out the heavy (in our case the Tera) dof's. We conjecture that in doing so a light (on the $\Lambda_T$ scale) scalar particle, denoted by $h$ in the following, remains in the spectrum which must then be included in the description of the low energy physics. This particle, that we propose to identify with the 125~GeV resonance discovered at LHC~\cite{Aad:2012tfa,Chatrchyan:2012xdj}, is interpreted within the present scheme as a composite $W^+W^-/ZZ$ state bound by Tera-particle exchanges. In sect.~\ref{sec:HIGGS} and in Appendix~\ref{sec:APPD} we provide well grounded arguments supporting this conjecture.

As a practical recipe to obtain the expression of the LEEL in terms of the effective fields, the second step of the procedure consists in formally crossing out from~(\ref{GW}) all the operators in which the heavy Tera-fields appear, while keeping all $d\leq 4$ terms invariant under $\chi_L\times\chi_R$ that can be constructed with the help of the remaining light dof's, i.e.\ $q,\ell,A_\mu,W_\mu,B_\mu,U$ and $h$. 

In doing so, one recognizes, as worked out in sect.~4 of~(I), that the unitarity of the mother theory~(\ref{LTOT}) imposes certain constraints on {\it a priori} unrelated parameters that make the $h$ and $U$ dependence of the resulting effective Lagrangian (with the exception of the scalar potential term) exactly of the form $\Phi=(k_v+h)U$. Although $k_v$ is not the vev of $\Phi$, but it is given by the relation $k_v=c_w\Lambda_T$, one gets from~(\ref{MASSES}), just like in the SM, the relation $k_v=M_W/g_w$ and the fact that the coupling of $h$ to fermion pairs is proportional to the fermion mass.

One thus sees that, at the leading order in the gauge couplings, the resulting expression of the $d=4$ LEEL of the model is formally equal to the SM Lagrangian with the exception, as we said above, of the scalar potential. The latter describes the self-interactions of the particle $h$, and has no reason to have the Mexican-hat shape of the SM potential. 

\section{The 125~GeV resonance and the SM}
\label{sec:HIGGS}

As we have anticipated in the previous section, since in the scenario we are advocating here (and following the idea first put forward in~\cite{Frezzotti:2014wja}) there appears to be no need for an elementary scalar particle of the kind introduced in refs.~\cite{Higgs:1964ia,Higgs:1964pj,Englert:1964et,Guralnik:1964eu} to give mass to fermions and EW gauge bosons, it is mandatory to provide a convincing interpretation for the 125~GeV resonance, identified at LHC by ATLAS~\cite{Aad:2012tfa} and CMS~\cite{Chatrchyan:2012xdj}. We propose to interpret this particle as a 
\beq
|h\rangle =|2\,W^+W^-+ZZ\rangle
\label{hWWZZ}
\eeq
composite state bound by exchanges of Tera-particles which are charged also under EW interactions. If this interpretation is correct, we expect, for instance, to see a pole in the $s$-channel of the $W^+W^-\!\to \!W^+W^-$ amplitude (as well as in $W^+W^-\!\to\! ZZ$ and $ZZ\!\to\! ZZ$ amplitudes) at $m_h^2$. In order to substantiate this interpretation we must be able to reliably compute the value of the $h$ binding energy, $E_{bind}$, and prove that
\beq
2M_W-E_{bind}=m_h \sim 125~{\mbox{GeV}}\, .
\label{WWEB}
\eeq

\subsection{A first try: a non-relativistic approach}
\label{sec:ANRA}

Although we do not have at the moment a rigorous argument supporting this conjecture, we show in Appendix~\ref{sec:APPD} that a single bound state of two weak bosons is formed, at least in the non-relativistic approximation, which is expected to be not too bad since $E_{bind}^2/4M_W^2 \ll 1$.\ In this limit we can provide an estimate of the $WW$ binding energy, based on a potential model description of the two-body $WW$ interaction, i.e.\ in terms of a particle of reduced mass $m = M_W/2$ in an attractive (approximately) square potential well of $depth \sim width \sim {\mbox{O}}(\Lambda_T)$. In Appendix~\ref{sec:APPD} we prove that the physical parameters characterizing Tera-dynamics are such that only one bound state can exist with a binding energy 
\beq
E_{bind}\sim c_{bind} \,g_w^\kappa M_W \, ,
\label{EBWW}
\eeq
where $c_{bind}$ is an O(1) computable constant and $\kappa$ a positive integer determined by the detailed dependence of the depth and width of the potential well on $g_w$. The $E_{bind}$ expression given by eq.~(\ref{EBWW}), which yields a value of the order of the $W$ boson mass itself, fits well with the phenomenological number, $2M_W-m_h\sim 160-125= 35$~GeV. 

\subsection{A better try: a Bethe--Salpeter-like approach}
\label{sec:ABT}

A theoretically more satisfactory approach to the calculation of the binding energy, $E_{bind}$, is to make recourse to the Bethe--Salpeter equation which, however, requires the knowledge of the effective $WW-WW$ coupling. The latter can be extracted from the four-point amputated correlator~\footnote{We do not display Lorentz and weak isospin indices in any of the formulae below, because at the crude level of this discussion we are unable to assess their role.}
\beqn
\hspace{-1.6cm}&&G_4 (p_1,p_2,p_3,p_4) =  \Big\langle \widehat W(p_1) \widehat W(p_2) \widehat W(p_3) \widehat W(p_4) \Big\rangle_{\rm amp} = g^4_w  \Big\langle \widehat J^L(p_1) \widehat J^L(p_2) \widehat J^L(p_3) \widehat J^L(p_4) \Big\rangle
\label{FWA}
\eeqn
where $\widehat W(p)$ and $\widehat J^L(p)$ are the Fourier transforms of $W(x)$ and $J^L(x)$, respectively. $G_4$ is dominated by the sum of Tera-strong exchanges. 

An estimate of the $WW$ binding energy, $E_{bind}$, due to Tera-exchanges,  can be obtained under the key assumption that the latter is (parametrically) small compared to $2M_W$. As we see from eq.~(\ref{FWA}), $E_{bind}$ is, indeed, expected to be proportional to $g_w^4$. In this situation the dominant contribution in the Bethe--Salpeter equation is the iteration of the kernel (see fig.~\ref{fig:DELTA}) 
\beqn
\hspace{-1.6cm}&&\Delta(s=4E^2,t=0)\, \delta(\vec p-\vec p')\delta(E-E')= g^4_w  \Big\langle \widehat J^L(\vec p, E) \widehat J^L(-\vec p, E) \widehat J^L(\vec p\,', E') \widehat J^L(-\vec p\,', E') \Big\rangle 
\label{GTPJ}
\eeqn
with attached almost on-shell $W$ legs. 
\begin{figure}[htbp]
\centerline{\includegraphics[scale=.35,angle=0]{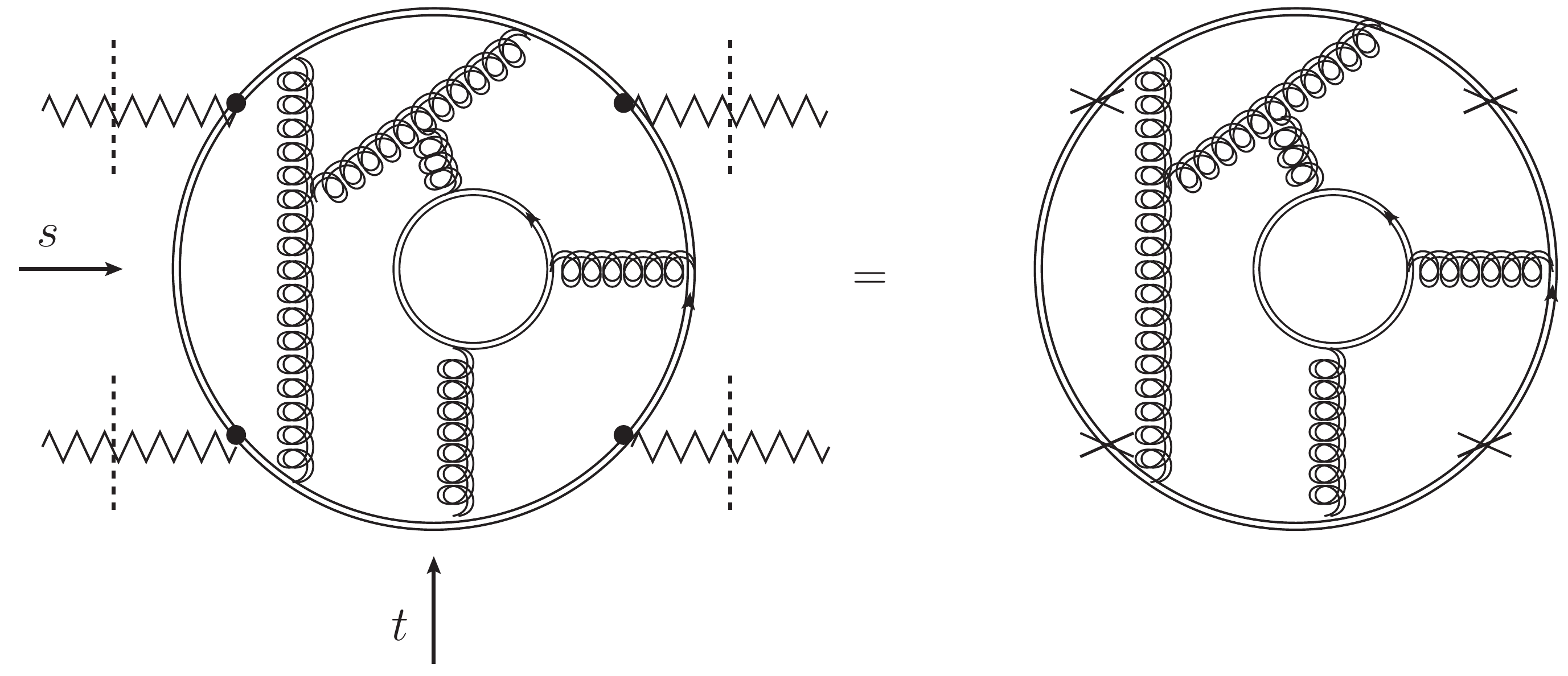}}
\caption{\small The amputated kernel $\Delta(s,t)$. The notations are like in fig.~\ref{fig:fig1}.} 
\label{fig:DELTA}
\end{figure}
As is well-known, $\Delta$ is directly related to the energy shift of the initial free state due to the interaction. In the case at hand, in fact, the Bethe--Salpeter iteration can be  cast in the approximated form (depicted in fig.~\ref{fig:figW}) 
\begin{figure}[htbp]
\centerline{\includegraphics[scale=.5,angle=0]{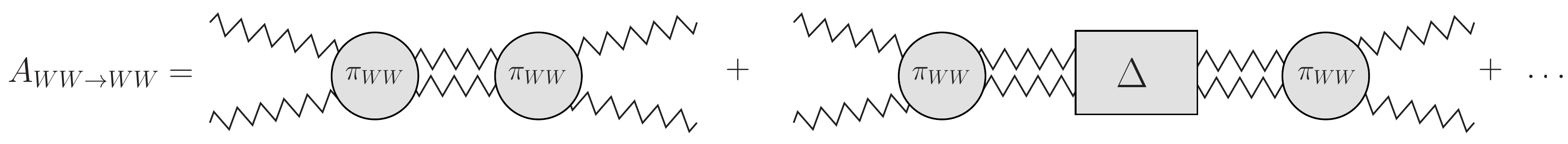}}
\caption{\small The iteration of eq.~(\ref{FWAS}).} 
\label{fig:figW}
\end{figure}
\beqn
\hspace{-.4cm}&&A_{WW\to WW}(s) ~\Rightarrow ~ \dfrac{(\pi_{WW}(s))^2}{s + 4M_W^2} \Big{[} 1 - \dfrac{\Delta(s,0)}{s + 4M_W^2} + \dots \Big{]} = \nn\\
\hspace{-.4cm}&&\qquad=\dfrac{(\pi_{WW}(s))^2}{s + 4M_W^2 + \Delta(s,0)} \stackrel{s \to s_{pole}}{\longrightarrow} \dfrac{g_w^2 M_W^2}{s+ m_h^2}\, ,
\label{FWAS}
\eeqn
where 
\beq
\pi_{WW}(s)\Big{|}_{s_{pole}} ={\mbox{O}}(g_w^2 \Lambda_T)={\mbox{O}}(g_w M_W)
\label{GA}
\eeq
is the price one needs to pay to have two $W$ sufficiently close to each other to be able to feel Tera-interactions. The resummation in eq.~(\ref{FWAS}) is very similar to the one that it is encountered in the Witten--Veneziano derivation of the $\eta'$ mass formula~\cite{Witten:1979vv,Veneziano:1979ec}. 

The obvious difference is that here the iterated, exchanged intermediate state is not the pion but the state of two loosely interacting $W$'s, whose invariant mass we approximate by simply setting mass$_{WW}^2\sim 4M_W^2$. The role of the topological susceptibility is here played by $\Delta$. From the second line of eq.~(\ref{FWAS}), one concludes that $\Delta(s,0)|_{s_{pole}}$ is the $WW$ binding energy. 

Owing to dimensional arguments and $g_w^2$ counting, we can write for the value of $\Delta(s)$ at the pole the parametric expression $\Delta(s,0)|_{s_{pole}}= c g_w^4 4 M_W^{\;2}$. The crucial assumption in this formula is the sign of $c$ that we conjecture to be negative, if it has to give rise to binding. We remark that the $g_w^2$ dependence of the residue at the pole in eq.~(\ref{FWAS}) correctly fits with the $g_w^2$ dependence of the SM $WW-WW$ scattering amplitude for which one gets
\beqn
A_4^{\rm SM}(s)\Big{|}_{pole} = \frac{g^2_{hWW}}{s+m^2_h}= \frac{4 g_w^2M^2_W}{s+m^2_h}\, .
\label{AEFF}
\eeqn
This consistency supports the correctness of the $g_w$ dependence of the coefficient $\pi_{WW}(s)|_{s_{pole}}$ displayed in eq.~(\ref{GA}).

In conclusion from~(\ref{FWAS}) one gets for the $h$-(mass)$^2$ 
\beq
m_h^2= 4M_W^2+c g_w^4 4 M_W^2\, .
\label{MHGG}
\eeq
Taking eq.~(\ref{MHGG}) at face value with $g_w=0.62$ and $c$ negative and ${\mbox{O}}(1)$, one obtains from 
\beq
m_h=2M_W\sqrt{1+cg_w^4} = 2M_W-E_{bind}
\label{EB}
\eeq
the interesting estimate 
\beq
E_{bind}=-c g_w^4 M_W [1+{\mbox{O}}(g_w^4)]\sim 12~{\mbox{GeV}}\, .
\label{EBMHIG}
\eeq
We point out that we have found a value of $E_{bind}$ of the order of the $W$ mass itself times powers of the weak coupling, like in the non-relativistic approximation described in  Appendix~\ref{sec:APPD}. The precise numerical magnitude of the estimate~(\ref{EBMHIG}) depends on  an ``at this moment unknown'' coefficient in eq.~(\ref{MHGG}) whose size and sign, however, could be extracted from the unquenched lattice QCD simulations of the four-point correlator~(\ref{GTPJ}).

Indeed, extending the discussion carried out in sect.~\ref{sec:LF}, it is interesting to note that, again neglecting weak and strong loop corrections, compared to Tera-strong effects, as well as the impact of Tera-leptons, $\Delta$ could be extracted from lattice simulations of the QCD type.\ The idea is that, pretending that the Tera-particles in the diagrams contributing to fig.~\ref{fig:DELTA} are quarks and gluons, one can obtain the physical value of $\Delta$ by rescaling the four-current amplitude evaluated in unquenched QCD-like simulations by the ratio $\Lambda_T/\Lambda_{QCD}$.\ In the approximation we are working, one can take for $J^L$ the naive, continuum-like expression of the {\it Left}-handed weak current. We can limit to simulate the amplitude $\Delta$ in QCD, and not in the much more costly toy-model introduced in ref.~\cite{Frezzotti:2014wja} (and considered in sect.~\ref{sec:LF} for the evaluation of the $W$ and Tera-quark mass), because NP effects do not appear to be relevant in the formation of the $h$ bound state. It will be enough to introduce by hand quark masses of appropriate values. 

\section{Conclusions and outlook}
\label{sec:CONC}

In this paper we have shown that it is possible to construct a bSMm in which all elementary particles get a dynamical mass from a unique NP field-theoretical feature, and not via the Higgs mechanism. 

This scenario was confirmed  in ref.~\cite{Capitani:2019syo} (see also refs.~\cite{Capitani:2017ucw,Capitani:2017trq,Capitani:2018jtx,Frezzotti:2018zsy} for preliminary results) by direct numerical investigations based on lattice simulations of the simplest field theoretical model displaying the NP mass generation mechanism identified in ref.~\cite{Frezzotti:2014wja}.

Dynamical mass terms for elementary fermions and EW bosons appear in the QEL of the critical (chiral invariant) theory as a sort of NP anomalies actually preventing the full restoration of the chiral $\tilde\chi_L\times\tilde\chi_R$ symmetry. 

In view of the parametric expression of these NP masses (given at the lowest loop order by the eqs.~(\ref{MASSES})), we have argued that to get the top quark and the $W/Z$ mass of the correct order of magnitude there must exists, besides SM particles, a super-strongly interacting sector gauge invariantly coupled to the SM dof's, in such a way that the RGI scale, $\Lambda_T$, of the whole theory encompassing SM matter and the particles of the new super-strongly interacting sector, is pushed to a value much larger than $\Lambda_{QCD}$, and of the order of a few TeV's. In this framework we get a natural interpretation of the EW scale as (a fraction of) the scale of the new interaction. 

Admittedly the approach we have developed is highly unusual. However, as shown in sect.~\ref{sec:SPCS}, using the $W$ mass as an input energy scale, from the formulae~(\ref{MASSES}) remarkably good estimates of the fermion masses of the heaviest family can be obtained, as well as reasonable values for $\Lambda_T$ and the mass of Tera-particles.

Obviously, lacking a Higgs boson (responsible for the existence of masses), the Higgs mass tuning problem does not even arise, while at the same time in the scenario we are advocating the issue of the metastability of the Universe needs to be reexamined~\cite{Cabibbo:1979ay,Ellis:2009tp,Alekhin:2012py,Salvio:2015cja}.

The 125~GeV scalar resonance detected at LHC, that suggestively we have denoted by $h$, is interpreted as a $W^*W^-/ZZ$ state bound by Tera-particle exchanges. To support this interpretation we have worked out a Bethe--Salpeter based estimate of its binding energy, $E_{bind}$. Despite the crudeness of our approximations (factors of 2 are out of control) we have found a surprisingly good value for $E_{bind}$. 

Upon integrating out the heavy Tera-dof's, one is left with the SM dof's plus a ``light'' (on the $\Lambda_T$ scale) $h$ scalar. We have shown in sect.~4 of~(I) that, including $h$ in the LEEL of the theory and enforcing $\chi_L\times\chi_R$ invariance as well as unitarity, we arrive to a low energy effective Lagrangian valid for momenta$^2\ll \Lambda_T^2$ that looks like the SM Lagrangian, except for scalar potential which, representing the $h$ self-interactions, has no reason to have the SM Mexican-hat shape.

We would like to conclude the discussion of this long investigation, started in ref..~\cite{Frezzotti:2014wja} and finalized in~(I) and in this paper, by saying that we have been able to develop the construction of an economic bSMm, in which elementary fermion and EW boson masses are dynamically determined and where we can, not only give a simple solution of the naturalness problem, but also understand the physical origin of the EW scale as being the scale of a new interaction. 

We also observe that, neglecting the presence of the supposedly existing new sector of heavy Tera-fermions in the calculation of the vacuum polarization contribution to the muon $g-2$ might be at the origin of the tension between theoretical expectations~\cite{Borsanyi:2020mff,Colangelo:2022jxc,Davier:2023cyp,Alexandrou:2022jlc} and the most recent measure of the muon anomalous magnetic moment~\cite{Muong-2:2023cdq}.

Checking the detailed predictions of models with NP mass generation against the wealth of the results of the existing experiments aimed at testing the validity of the SM and searching for New Physics, is obviously of paramount importance. We anticipate that this task is not going be easy, as most of the features of the kind of models we are advocating here are of NP origin. Certainly, the direct production of Tera-hadrons in accelerator experiments would be an unmistakable sign of New Physics.

\renewcommand{\thesection}{A} 
\section{Hypercharge assignments}  
\label{sec:APPA}

In this Appendix for completeness we recall the standard hypercharge assignment of SM particles (see Table~\ref{tab:standard}) and the peculiar one of Tera-particles chosen so as to give U(1)$_Y$, SU(2)$_L$ and SU($N_c=3$) gauge coupling unification~\cite{Frezzotti:2016bes} (see Table~\ref{tab:alternative}). Notice that consistently with eq.~(\ref{DER}) we have used the definition of hypercharge given by the relation ${\cal Q}=T_3+Y$, like in  ref.~\cite{PESK}. As we remarked in~\cite{Frezzotti:2016bes}, the Tera-fermion hypercharges are uniquely fixed by the requirement of anomaly cancellation. 
\begin{table}[htbp]
\begin{center}
\begin{tabular}{|l|l|}
\hline 
\vspace{.1cm}
$q$ & $\ell$   \\
\hline
\hline
\vspace{.1cm}
 $y_{u_L}=\frac{2}{3}-\frac{1}{2}=\frac{1}{6}$ & $y_{\nu_L}=0-\frac{1}{2}=-\frac{1}{2}$
\\
\hline
 \vspace{.1cm}  
 $y_{u_R}=\frac{2}{3}-0=\frac{2}{3}$ & $y_{\nu_R}=0$ \\
\hline
 \vspace{.1cm}  
 $y_{d_L}=-\frac{1}{3}+\frac{1}{2}=\frac{1}{6}$ & $y_{el_L}=-1+\frac{1}{2}=-\frac{1}{2}$   \\
\hline
 \vspace{.1cm}  
 $y_{d_R}=-\frac{1}{3}-0=-\frac{1}{3}$ & $y_{el_R}=-1-0=-1$   \\
\hline
\hline
\vspace{.1cm}  
$\sum y^2_q =\frac{22}{36}$ & $\sum y^2_\ell =\frac{3}{2}$\\
\hline
\end{tabular}
\caption{\small{Hypercharges of SM fermions}}
\label{tab:standard}
\end{center}
\end{table}
\begin{table}[htbp]
\begin{center}
\begin{tabular}{|l|l|}
\hline
\vspace{.1cm}
$Q$  & $L$ 
 \\
\hline
\hline
\vspace{.1cm}
$y_{U_L}=\frac{1}{2}-\frac{1}{2}=0$ & $y_{N_L}=\frac{1}{2}-\frac{1}{2}=0$   \\
\hline
\vspace{.1cm}
$y_{U_R}=\frac{1}{2}-0=\frac{1}{2}$ & $y_{N_R}=\frac{1}{2}-0=\frac{1}{2}$   \\
\hline
\vspace{.1cm}
$y_{D_L}=-\frac{1}{2}+\frac{1}{2}=0$ & $y_{L_L}=-\frac{1}{2}+\frac{1}{2}=0$ \\
\hline
\vspace{.1cm}
$y_{D_R}=-\frac{1}{2}-0=-\frac{1}{2}$ & $y_{L_R}=-\frac{1}{2}-0=-\frac{1}{2}$ \\
\hline
\hline
\vspace{.1cm}
$\sum y^2_Q =\frac{1}{2}$ & $\sum y^2_L =\frac{1}{2}$ \\
\hline
\end{tabular}
\caption{\small{Tera-particle hypercharges}}
\label{tab:alternative}
\end{center}
\end{table}

Two peculiar features of the Tera-hypercharge assignment of Table~\ref{tab:alternative} are worth noticing 1) Tera-fermions have half-integer electric charges, 2) {\it Left}-handed components have vanishing hypercharge. An ``amusing'' consequence of this fact is that {\it Left}-handed components of Tera-fermions do not couple to $B$ but only to $W$, while the opposite is true for the {\it Right}-handed components.

\renewcommand{\thesection}{B} 
\section{A diagrammatic interpretation of the $d=6$ NP O($b^2$) operators}  
\label{sec:APPB}

It is instructive to try to {\it heuristically} explain how physically the NP O($b^2 \Lambda_T$) Symanzik operators~(\ref{OTT})--(\ref{OBBL}) arise in the NG phase of the theory. They emerge as the result of an interplay between the residual O($b^2$) explicit violations of the (recovered) $\tilde\chi_L \times \tilde\chi_R$ chiral symmetry and the effects of the dynamical spontaneous breaking of the latter occurring in the NG phase of the critical model, in turn triggered by O($b^2 v$) chiral breaking terms. 

For the development of the argument we must imagine to be dealing with the fundamental Lagrangian~(\ref{LTOT}) in the NG phase in a perturbative loop expansion. The key conjecture consists in assuming that the formally O($b^{2}$) chiral breaking term of the Tera-fermions (Tera-quark or Tera-lepton) propagator, $\Pi^{\tilde\chi{\rm-br}}$, get modified by NP effects~\footnote{A conceptually similar line of arguments based on an extension of the Banks--Casher formula~\cite{Banks:1979yr} for the expression of the eigenvalue density of the Dirac operator was elaborated in ref.~\cite{Frezzotti:2014wja}, showing that indeed in Wilson lattice QCD an O($\Lambda_{QCD}$) term gets NP-ly generated in the quark critical mass.}. 

More in detail, the trigger of the whole NP phenomenon should be identified with the contribution to the Tera-fermion propagator, coming from the chiral breaking part of the Tera-Wilson-like term with the scalar set at $v$, that we {\it conjecture} appears multiplied by the typical NP factor $\exp (- [2\beta_{T0}\,g_T^2(\rho_T v)]^{-1})$. In the exponent $\beta_{T0}$ is the first coefficient of the $\beta$-function of Tera-strong interactions with the running Tera-gauge coupling constant computed at the scale $\rho_T v\to\rho_Q v$ in the case of Tera-quarks or $\rho_T v\to\rho_L v$ in the case of Tera-leptons. The choice of these scales is quite natural because, as we have shown in the evaluation of the self-energy diagrams of figs.~\ref{fig:fig1} and~\ref{fig:fig11} (see also similar calculations in~\cite{Frezzotti:2014wja} and~(I)), the finite NP masses come from the region of phase space where all the loop momenta are O($b^{-1}$). In such a kinematical situation the order of magnitude of the chiral symmetry breaking Tera-quark Wilson-like term is $\rho_Q v b^2 k^2 \sim \rho_Q v$ and that of the Tera-lepton Wilson-like term $\rho_L v b^2 k^2 \sim \rho_L v$. Putting the various bits together we get
\beq
\Pi^{\tilde\chi{\rm -br}}_{Q/L\,mn}(k)= \frac{1}{k^2} \rho_{Q/L} v b^2 k^2 \exp \Big{[}\!- \frac{1}{2\beta_{T0}g_T^2(\rho_{Q/L} v)}\Big{]}\,\delta_{mn} = b^2 s_{\rho_{Q/L}}\Lambda_T\delta_{mn}\, ,
\label{CHUBRPROP}
\eeq
where $m,n$ are spin indices and $s_{\rho_{Q/L}}$ denotes the sign of $\rho_{Q/L}$ which can always be taken to be positive. In a diagram we will denote this ``basic'' NP effect on the Tera-fermion propagator by a double line with a cross. A few observations are in order here. 

1) NP effects of the kind described above also occur in the chiral breaking part of the quark propagator. However, they will contribute negligibly to O($b^2$) Symanzik operators as the quark analog of eq.~(\ref{CHUBRPROP}) will be scaled down by a factor $\Lambda_{QCD}/\Lambda_T$.

2) At the lowest loop order $\Pi^{\tilde\chi{\rm-br}}$ is a $\delta$-function in configuration space.

3) As we see from eq.~(\ref{CHUBRPROP}), the chiral breaking terms in the Tera-quark and Tera-lepton propagators are equal. 

4) In building more complicated diagrams one should keep in mind that the expression in the r.h.s.\ of eq.~(\ref{CHUBRPROP}), wherever it enters, needs in the end to be multiplied by $U$ in order to make things properly $\chi_L\times\chi_R$ invariant.

If we now consider a loop involving the propagator~(\ref{CHUBRPROP}) and a Tera-gluon line going from a ``standard'' Tera-strong interaction vertex (black dot) to a Wilson-like vertex (gray square), as shown in the two diagrams of fig.~\ref{fig:fig12}~\footnote{This sort of asymmetry between the two vertices is necessary in order for the completeness relation of the eigenfunctions of the full fermion Dirac operator not to wash out the NP effect~\cite{Frezzotti:2014wja}.}, we get contributions of O($b^2\Lambda_T\alpha_T$) that can be described by the two NP Symanzik operators~(\ref{OTT}) and~(\ref{OLL}). When inserted in more complicated diagrams, like for instance the self-energy diagrams in figs.~\ref{fig:fig1}, for simplicity they will be indicated with a blob with external Tera-fermion--scalar--anti-Tera-fermion legs.

In fig.~\ref{fig:fig13} we display the lowest loop order diagrams that, upon closing with a Tera-fermion line the chiral breaking piece of the Tera-propagator~(\ref{CHUBRPROP}), give rise to the operators from~(\ref{OAA}) to~(\ref{OBBL}) that are O($b^2\Lambda_T g^2_p$) with $p=s,T,Y$ in the upper, medium and lower panel, respectively. Similarly to what we did before, when inserted in more complicated diagrams, like for instance the self-energy diagrams in fig.~\ref{fig:fig1} or fig.~\ref{fig:fig11}, for simplicity the operators from~(\ref{OAA}) to~(\ref{OBBL}) will be indicated with a blob with external 
gauge-field--scalar--gauge-field legs.

One might have noticed that in the operators of fig.~\ref{fig:fig12} we have parametrized the gauge coupling dependence by factors $\alpha$, while in the operators of fig.~\ref{fig:fig13} by factors $g^2$. The reason for this difference is that in the diagrams of fig.~\ref{fig:fig12} the Tera-gluon closes a loop, thus supposedly giving raise to a $g^2_T/4\pi=\alpha_T$ factor, while in the diagrams of fig.~\ref{fig:fig13} the gauge fields are external legs coupled by two $g_p$ factors with $p=s,T,Y$ according to the diagrams in the panels we refer to. What we have described is what we have called ``optimal'' choice in sect.~\ref{sec:MASS} and in sect.~3.4 of~(I).

We end by observing that the appearance of the modulus of the scalar field in the expressions of these Symanzik operators is due to the cancellation of the $\Phi$ phase operated by the $U$ factor (not shown) that multiplies the expression~(\ref{CHUBRPROP}), as we remarked in 4) above.
\begin{figure}[hpt]
\centerline{\includegraphics[scale=0.5]{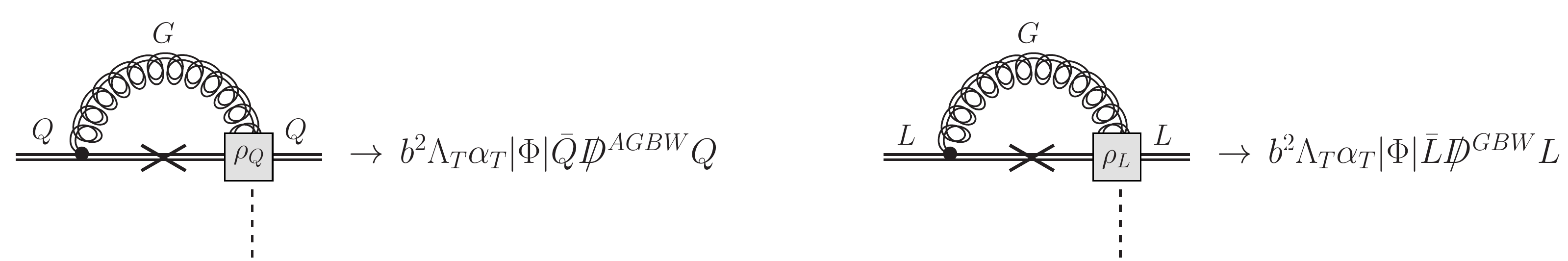}}
\caption{\small{First panel: representative lowest loop order diagrams yielding the dynamically generated operators~(\ref{OTT}) and~(\ref{OLL}), respectively. Double lines labelled by $Q$ and $L$ represent Tera-quark and Tera-leptons, respectively. Curly double lines labelled by $G$ represent Tera-gluons. Dotted line is $|\Phi|$.}}
\label{fig:fig12}
\end{figure}
\begin{figure}[hpt]
\centerline{\includegraphics[scale=0.5]{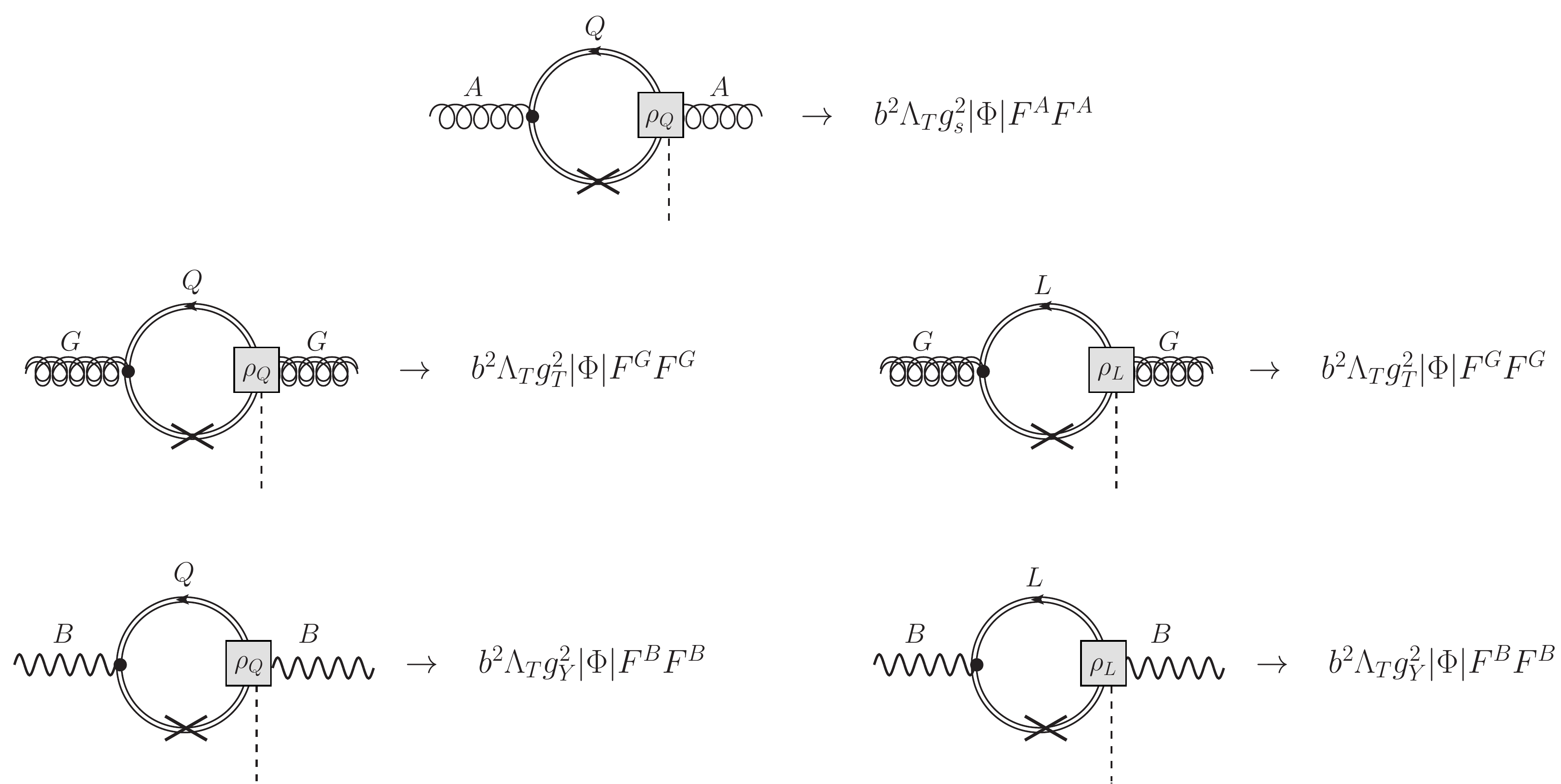}}
\caption{\small{Typical lowest loop order diagrams yielding the dynamically generated operators from~(\ref{OAA}) to~(\ref{OBBL}). Curly single lines are gluons, wiggly lines are $W$ and wavy lines $B$ bosons. For the rest, notations are as in fig.~\ref{fig:fig12}.}}
\label{fig:fig13}
\end{figure}

\renewcommand{\thesection}{C} 
\section{The $\rho$ dependence of physical quantities}  
\label{sec:APPC}

We prove in this Appendix that in the presence of weak interactions the physics of the critical theory only depends on the ratios of the $\rho$ coefficients multiplying the various Wilson-like terms of the Lagrangian. We want to stress that, as we are going to prove, it is just the presence of weak interactions that makes the observables of the critical theory only functions of $\rho$ ratios, thus in a sense ``softening'' the dependence of physical quantities on the (at the moment arbitrary) $\rho$ parameters.

We start by considering the simpler model (with neither leptonic matter nor hypercharge), studied in~(I) whose Lagrangian for completeness we report here 
\beqn
\hspace{-.9cm}&&{\cal L}(q,Q;A,G,W;\Phi)\!= \label{SULLWQ}\\
\hspace{-.9cm}&&\qquad={\cal L}_{K}(q,Q;A,G,W;\Phi)\!+\!{\cal V}(\Phi)\!+\!{\cal L}_{Yuk}(q,Q;\Phi)\!+\!{\cal L}_{Wil}(q,Q;A,G,W;\Phi) \nn
\eeqn
\beqn
\hspace{-.9cm}&&\,\,\bullet\,\,{\cal L}_{K}(q,Q;A,G,W;\Phi)= \frac{1}{4}\Big{(}F^A\cdot F^A+F^G\cdot F^G+F^W\cdot F^W\Big{)}+\label{LKINWQ}\\
\hspace{-.9cm}&&\quad+\big{(}\bar q_L\,/\!\!\!\!{\cal D}^{AW} q_L\!+\!\bar q_R\,/\!\!\!\!{\cal D}^{A} q_R\big{)}\!+\!\big{(}\bar Q_L\,/\!\!\!\!{\cal D}^{AGW} Q_L\!+\!\bar Q_R\,/\!\!\!\!{\cal D}^{AG} Q_R\big{)}\!+\!\frac{k_b}{2}{\tr}\big{[}({\cal D}\,^{W}_\mu \Phi)^\dagger{\cal D}^W_\mu\Phi\big{]}\nn\\
\hspace{-.9cm}&&\,\,\bullet\,\,{\cal V}(\Phi)= \frac{\mu_0^2}{2}k_b{\tr}\big{[}\Phi^\dagger\Phi\big{]}+\frac{\lambda_0}{4}\big{(}k_b{\tr}\big{[}\Phi^\dagger\Phi\big{]}\big{)}^2\label{VWQ}\\
\hspace{-.9cm}&&\,\,\bullet\,\,{\cal L}_{Yuk}(q,Q;\Phi)=\eta_q\,\big{(} \bar q_L\Phi\, q_R+\bar q_R \Phi^\dagger q_L\big{)} + \eta_Q\,\big{(} \bar Q_L\Phi\, Q_R+\bar Q_R \Phi^\dagger Q_L\big{)}\label{LYUKWQ} \\
\hspace{-.9cm}&&\,\,\bullet\,\,{\cal L}_{Wil}(q,Q;A,G,W;\Phi)= \frac{b^2}{2}{\rho_q}\,\big{(}\bar q_L{\overleftarrow {\cal D}}\,^{AW}_\mu\Phi {\cal D}^A_\mu q_R+\bar q_R \overleftarrow{\cal D}\,^A_\mu \Phi^\dagger {\cal D}^{AW}_\mu q_L\big{)}+\nn\\
\hspace{-.9cm}&&\quad+\frac{b^2}{2}{\rho_Q}\,\big{(}\bar Q_L{\overleftarrow {\cal D}}\,^{AGW}_\mu\Phi {\cal D}^{AG}_\mu Q_R+\bar Q_R \overleftarrow{\cal D}\,^{AG}_\mu \Phi^\dagger {\cal D}^{AGW}_\mu Q_L\big{)}\, .\label{LWILWQ}
\eeqn
We will prove that physics only depends on the ratio $r_{Qq}=\rho_Q/\rho_q$ and not on $\rho_Q$ and $\rho_q$ separately. We will extend the argument to an arbitrary number of fermion species in subsect.~\ref{sec:MTTF}.

To prove this statement it is convenient to first work out what happens when one performs in the Lagrangian~(\ref{SULLWQ}) the field rescaling $\widehat\Phi=\xi\Phi$ with a constant $\xi$. The kinetic Lagrangian terms of quarks, Tera-quarks and gauge fields are obviously left invariant. For the rest one gets 
\beqn
&&{\cal L}_{kin}(\Phi;W)=\widehat{\cal L}_{kin}(\widehat\Phi;W)=\frac{k_b}{2\xi^2}{\tr}\big{[}({\cal D}\,^W_\mu \widehat\Phi)^\dagger{\cal D}^W_\mu\widehat\Phi\big{]}\, , \label{SLKINP}\\
&&{\cal V}(\Phi)=\widehat{\cal V}(\widehat\Phi)= \frac{\mu_0^2}{2}\frac{k_b}{\xi^2}{\tr}\big{[}\widehat\Phi^\dagger\widehat\Phi\big{]}+\frac{\lambda_0}{4}\Big{(}\frac{k_b}{\xi^2}{\tr}\big{[}\widehat\Phi^\dagger\widehat\Phi\big{]}\Big{)}^2\, ,\label{SLV2}\\
&& {\cal L}_{Yuk}(q,Q;\Phi)=\widehat{\cal L}_{Yuk}(q,Q;\widehat\Phi) =\nn\\
&&\quad =\frac{\eta_q}{\xi}\,\big{(} \bar q_L\widehat\Phi\, q_R+\bar q_R\widehat \Phi^\dagger q_L\big{)} + \frac{\eta_Q}{\xi}\,\big{(} \bar Q_L\widehat\Phi\, Q_R+\bar Q_R \widehat\Phi^\dagger Q_L\big{)}\, ,\label{SLLYUK}\\
&&{\cal L}_{Wil}(q,Q;\Phi;A,G,W)=\widehat{\cal L}_{Wil}(q,Q;\widehat\Phi;A,G,W)=\nn\\
&&\quad = \frac{b^2}{2}\frac{\rho_q}{\xi}\,\big{(}\bar q_L{\overleftarrow {\cal D}}\,^{AW}_\mu\widehat\Phi {\cal D}^A_\mu q_R+\bar q_R \overleftarrow{\cal D}\,^A_\mu \widehat\Phi^\dagger {\cal D}^{AW}_\mu q_L\big{)}+\nn\\
&&\quad +\frac{b^2}{2}\frac{\rho_Q}{\xi}\,\big{(}\bar Q_L{\overleftarrow {\cal D}}\,^{AGW}_\mu\widehat\Phi {\cal D}^{AG}_\mu Q_R+\bar Q_R \overleftarrow{\cal D}\,^{AG}_\mu \widehat\Phi^\dagger {\cal D}^{AGW}_\mu Q_L\big{)} \, .\label{SLWIL}
\eeqn
The r.h.s.\ of the equations from~(\ref{SLKINP}) to~(\ref{SLWIL}) have the same form as the corresponding terms in the Lagrangian~(\ref{SULLWQ}), except for the replacements $k_{b}\to k_{b}/\xi^2$, $\eta_{q;Q}\to\eta_{q;Q}/\xi$, $\rho_{q;Q}\to\rho_{q;Q}/\xi$ and obviously $\Phi\to\widehat\Phi$. Consequently (with simplified notations) 
\beqn
\hspace{-1.4cm}&&{\mbox{if}}\quad\frac{b^2}{2}\rho_q \,\big{(}\bar q_L{\overleftarrow {\cal D}}\,^{AW}_\mu\Phi {\cal D}^A_\mu q_R+ {\mbox{hc}}\big{)} \,\,{\mbox{mixes with}} \nn\\
\hspace{-1.4cm}&&\qquad \bar \eta_q(\eta_q,\eta_Q,\rho_q,\rho_Q;v^2,\lambda_0) \,\big{(} \bar q_L\Phi\, q_R+{\mbox{hc}}\big{)+\ldots} \, ,\label{MIX1}
\eeqn
\beqn
\hspace{-1.4cm}&&\mbox{then}\quad\frac{b^2}{2}\frac{\rho_q}{\xi}\,\big{(}\bar q_L{\overleftarrow {\cal D}}\,^{AW}_\mu\widehat\Phi {\cal D}^A_\mu q_R+ {\mbox{hc}}\big{)} \,\,{\mbox{mixes with}}\nn\\
\hspace{-1.4cm}&&\qquad \quad \bar\eta_q\Big{(}\frac{\eta_q}{\xi}, \frac{\eta_Q}{\xi}, \frac{\rho_q}{\xi},\frac{\rho_Q}{\xi};v^2,\lambda_0\Big{)}\,\big{(} \bar q_L\widehat\Phi\, q_R+{\mbox{hc}}\big{)} +\ldots\, . \label{MIX2}
\eeqn
As a result the critical condition for $\eta_q$ 
\beqn
\eta_q= \bar \eta_q(\{g\};\eta_q,\eta_Q,\rho_q,\rho_Q;v^2,\lambda_0) 
\label{EQCR1}
\eeqn
turns into 
\beqn
\frac{\eta_q}{\xi}=\bar \eta_q \Big{(}\{g\};\frac{\eta_q}{\xi}, \frac{\eta_Q}{\xi}, \frac{\rho_q}{\xi},\frac{\rho_Q}{\xi};v^2,\lambda_0\Big{)} \, .
\label{EQCR2}
\eeqn
Similar equations are valid for $\eta_Q$ and $k_b$, so that the structure of the solution of the criticality equations will be
\beqn
&&\frac{\eta_{q\,cr}}{\xi} = f_{\eta_q}\Big{(}\{g\};\frac{\rho_q}{\xi},\frac{\rho_Q}{\xi};v^2,\lambda_0\Big{)}\, ,\label{SSOLDEPV1}\\
&&\frac{\eta_{Q\,cr}}{\xi} =f_{\eta_Q}\Big{(}\{g\};\frac{\rho_q}{\xi},\frac{\rho_Q}{\xi};v^2,\lambda_0\Big{)}\, ,\label{SSOLDEPV2}\\
&&\frac{k_{b\,cr}}{\xi^2} =f_{k_b}\Big{(}\{g\};\frac{\rho_q}{\xi},\frac{\rho_Q}{\xi};v^2,\lambda_0\Big{)}\label{SSOLDEPV3}
\eeqn
with $f_{\eta_q}$, $f_{\eta_Q}$ and $f_{k_b}$ functions of the arguments displayed above. 

\subsection{The canonical critical Lagrangian}

We now perform the canonical rescaling $\sqrt{k_{b\,cr}}\,\Phi\!=\!\widetilde\Phi$, in the critical Lagrangian~(\ref{SULLWQ}), obtaining (we only display the terms affected by the rescaling)
\beqn
\hspace{-1.2cm}&&\widetilde{\cal L}_{kin}(\widetilde\Phi;W)=\frac{1}{2}{\tr}\big{[}({\cal D}\,^W_\mu \widetilde\Phi)^\dagger{\cal D}^W_\mu\widetilde\Phi\big{]}\, , \label{SLKINPT}\\
\hspace{-1.2cm}&&\widetilde{\cal V}(\widetilde\Phi)= \frac{\mu_0^2}{2}{\tr}\big{[}\widetilde\Phi^\dagger\widetilde\Phi\big{]}+\frac{\lambda_0}{4}\big{(}{\tr}\big{[}\widetilde\Phi^\dagger\widetilde\Phi\big{]}\big{)}^2 \, ,\label{SLV2T}\\
\hspace{-1.2cm}&& \widetilde{\cal L}_{Yuk}(q,Q;\widetilde\Phi) =\nn\\
\hspace{-1.2cm}&&\quad =\frac{\eta_{q\,cr}}{\sqrt{k_{b\,cr}}}\,\big{(} \bar q_L\widetilde\Phi\, q_R+\bar q_R\widetilde\Phi^\dagger q_L\big{)} + \frac{\eta_{Q\,cr}}{\sqrt{k_{b\,cr}}}\,\big{(} \bar Q_L\widetilde\Phi\, Q_R+\bar Q_R \widetilde\Phi^\dagger Q_L\big{)}\, ,\label{SLYUKT}\\
\hspace{-1.2cm}&&\widetilde{\cal L}_{Wil}(q,Q;\widetilde\Phi;A,G,W)=\nn\\
\hspace{-1.2cm}&&\quad = \frac{b^2}{2}\frac{\rho_q}{\sqrt{k_{b\,cr}}}\,\big{(}\bar q_L{\overleftarrow {\cal D}}\,^{AW}_\mu\widetilde\Phi {\cal D}^A_\mu q_R+\bar q_R \overleftarrow{\cal D}\,^A_\mu \widetilde\Phi^\dagger {\cal D}^{AW}_\mu q_L\big{)}+\nn\\
\hspace{-1.2cm}&&\quad +\frac{b^2}{2}\frac{\rho_Q}{\sqrt{k_{b\,cr}}}\,\big{(}\bar Q_L{\overleftarrow {\cal D}}\,^{AGW}_\mu\widetilde\Phi {\cal D}^{AG}_\mu Q_R+\bar Q_R \overleftarrow{\cal D}\,^{AG}_\mu \widetilde\Phi^\dagger {\cal D}^{AGW}_\mu Q_L\big{)} \, .\label{SLWILT}
\eeqn
According to eqs.~(\ref{SLKINPT})--(\ref{SLWILT}) we see that physics only depends on the ratios $\eta_{q\,cr}/\sqrt{k_{b\,cr}}$, $\eta_{Q\,cr}/\sqrt{k_{b\,cr}}$, $\rho_{q}/\sqrt{k_{b\,cr}}$ and $\rho_{Q}/\sqrt{k_{b\,cr}}$. From eqs.~(\ref{SSOLDEPV1})--(\ref{SSOLDEPV3}) we conclude that these ratios  are functions of $\rho_q/\xi$ and $\rho_Q/\xi$ (besides the gauge couplings and possibly $\lambda_0$, but for dimensional reasons not of $v^2$). Since the integral that defines the (generating functional of the) theory is $\xi$ rescaling independent, taking $\xi\!=\!\rho_q$ (or $\xi\!=\!\rho_Q$) immediately proves our Theorem.

\subsubsection{Observations}
\label{sec:OBSER}

1) We note that, if we have only one species of fermions, hence only one kind of Yukawa and Wilson-like terms in the presence of weak interactions, then from the above argument it follows that the NP-ly generated masses do not depend on $\rho$.

2) One can show that up to radiative corrections, when $\rho_q$ and $\rho_Q$ vary from $0^+$ to $\infty$, the ratios $\rho_{q}/\sqrt{k_{b\,cr}}$ and $\rho_{Q}/\sqrt{k_{b\,cr}}$ vary in the open intervals $(0, 1/\sqrt{c_q \nu_q})$ and $(0, 1/\sqrt{C_Q\nu_Q})$, respectively, with $c_q$ and $C_Q$ computable coefficients and the integers $\nu_q$ and $\nu_Q$ the number of the fermions running in the loops that determine $k_{b\,cr}$ (see fig.~\ref{fig:fig22}).\ By $\nu_h$ we mean the product of the dimensions of the representations of the internal symmetry groups (weak isospin, colour and/or Tera-colour) to which each fermion belongs. The proof is given in the next subsection, where we consider the more general case of the Lagrangian~(\ref{LTOT}).

\subsection{The case of more than two fermion species}
\label{sec:MTTF}

We extend the previous analysis by assuming that we are dealing with the realistic case of $f=q,Q,\ell,L$ species of fermions. In the fundamental Lagrangian (see eq.~(\ref{LTOT})) there is a $d=6$ Wilson-like term for each species, scaled by the parameter $\rho_f$ (which without loss of generality can all be taken positive). Extending the argument developed in Appendix~C of ref.~(I), one can write for the critical value of $k_b$ the expression (see fig.~\ref{fig:fig22})
\beq
k_{b\,cr}(\rho)=\sum_{f=q,Q,\ell,L} k^{(1)}_{b\,f} \nu_f \rho_f^2\Big{[}1+\ldots\Big{]}\, ,
\label{KBCR}
\eeq
where the coefficients $k^{(1)}_{b\,f} ={\mbox{O}}(1)$ are computable quantities and the dots in the square parenthesis represent small perturbative, higher loop corrections. As we said, for each $f$ the number $\nu_f$ is the multiplicity of corresponding fermion species contributing to the loop diagrams in fig.~\ref{fig:fig22}. Naturally if there is more than one SM family,, $k_{b\,cr}(\rho)$ will also depend on their number.\ In the general case of $N_f$ SM families one has $\nu_q=2N_f N_c$, $\nu_Q= 2N_T N_c$, $\nu_\ell=2N_f$ and $\nu_L=2N_T$. The presence of the factor 2 is due to the fact that fermions come in weak isospin doublets, both members of which contribute to the diagrams of fig.~\ref{fig:fig22}. 
\begin{figure}[htbp]
\centerline{\includegraphics[scale=0.35,angle=0]{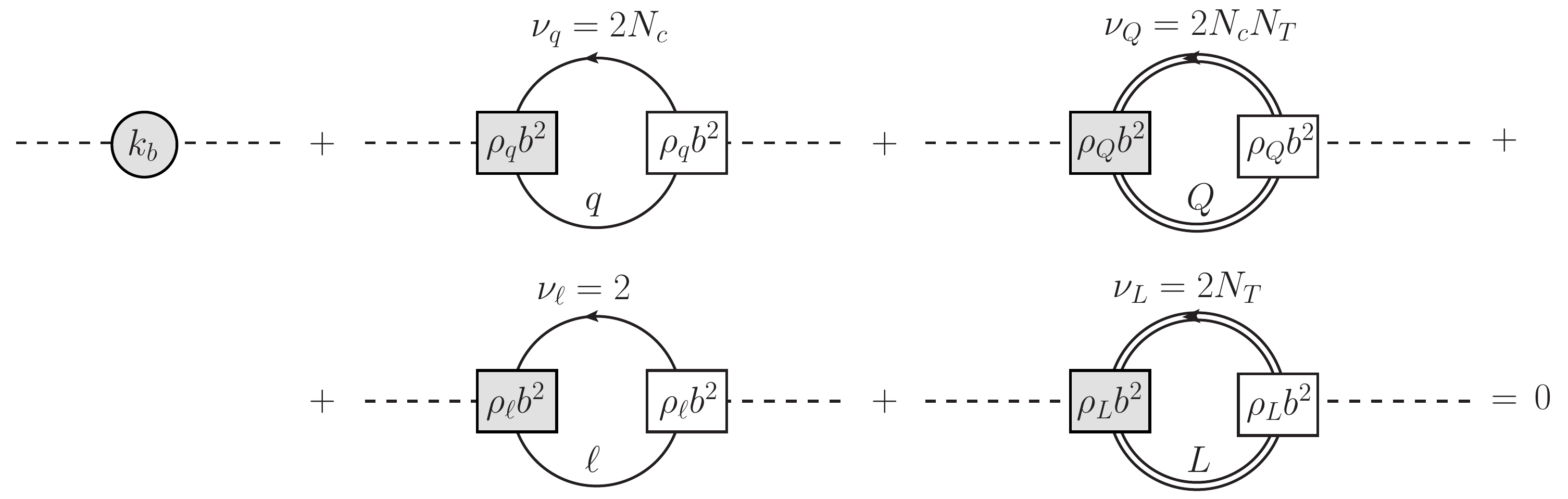}}
\caption{\small{The condition implied by the tuning condition determining $k_{b\,cr}$ (leading to the cancellation of the scalar kinetic term) at the lowest loop order in the Wigner phase. This figure is a generalization of fig.~6 of~(I) extended to include leptons and Tera-leptons. The integers $\nu_q$, $\nu_Q$ $\nu_\ell$ and $\nu_L$ are the multiplicities of quarks, Tera-quarks, leptons and Tera-leptons running in the loops with $N_c$ and $N_T$ the number of colours and Tera-colours, respectively. The multiplicities in the figure refer to the one-family ($N_f=1$) case. The grey disc represents the insertion of the scalar kinetic term. The grey box represents the operator the scalar kinetic term mixes with. The empty box represents the insertion of Wilson-like vertices from the Lagrangian necessary to close the loop. The rest of the notations is as in fig.~\ref{fig:fig1}.}}
\label{fig:fig22}
\end{figure}

We want to prove that, up to perturbative, higher loop  corrections, the rescaled parameters 
\beq
\widehat\rho_f=\dfrac{\rho_f}{\sqrt{k_{b\,cr}}}=\dfrac{\rho_f}{\Big{(}\sum_{h=q,Q,\ell,L} k^{(1)}_{b\,h}  \nu_h \rho_h^2\Big{)}^{1/2}}\Big{[}1+\ldots\Big{]}\, , \quad f=q,Q,\ell,L \, ,
\label{HATR}
\eeq
upon which the physics of the model effectively depends, are all bound from above when we let $\rho_f$ vary between $0^+$ and $\infty$.\ This feature is an immediate consequence of the criticality condition 
\beq
\sum_{f=q,Q,\ell,L} k^{(1)}_{b\,f} \nu_f \widehat \rho_f^2=1+\ldots \, ,
\label{SR}
\eeq
where dots represent perturbative, higher loop  corrections. We see that, ignoring the latter (which by the way partly compensate in the ratios~(\ref{HATR})), eq.~(\ref{SR}) entails the bounds
\beq
0<\widehat\rho_f<\frac{1}{(k^{(1)}_{b\,f} \nu_f)^{1/2}} \, ,\qquad f=q,Q,\ell,L\, .
\label{HRK}
\eeq
We end with a few remarks. 

1) As we have explicitly indicated in the eqs.~(\ref{HRK}), every $\rho_f$ (and consequently every $\widehat \rho_f$) needs to be taken strictly non-vanishing for the corresponding Wilson-like term to be able to trigger the NP mechanism that generates a mass for the fermion $f$. 

2) Concerning the overall dependence of the self-energy diagrams from $N_c$, $N_T$, it can be shown that kinematically multiplicities like either $N_T^2$ or $N_c N_T$ accompany $\rho^2$ factors, thus making the product [${\widehat \rho_f}^{\,2} \times$\,multiplicities] independent of $N_c$ and $N_T$ in the limit of large number of colours = number of Tera-colours (at fixed gauge couplings). This argument is to say that self-energy diagrams have a mild dependence on $N_c$ and $N_T$. 

In the limit in which we retain only the leading dependence on $N_c$ and $N_T$, the NP self-energy diagrams generated by the Lagrangian~(\ref{LTOT}) in which all Wilson-like terms are $d=6$ operators, have at 1-loop the parametric expression 
\beqn
&&m_q=z_q\frac{\rho_q}{\rho_Q}\alpha_s g^2_s\Lambda_T\label{MQUARK}\\
&&m_\ell=z_\ell\frac{\rho_\ell}{\rho_Q}\alpha_Y g^2_Y\Lambda_T\label{MELL}\\
&&m_Q=z_Q\alpha_T g^2_T \Lambda_T\label{MTQUARK}\\
&&m_L=\Big{(}z_{1L}\frac{\rho_L}{\rho_Q}+z_{2L}\frac{\rho_L^2}{\rho_Q^2}  \Big{)}\,\alpha_T g^2_T\Lambda_T\label{MTELL}\\
&&M_W=z_W\alpha_T g_w \Lambda_T\label{MWW}
\eeqn
with $z_f$ suitable finite numerical constants that do not depend on the $\rho$'s. Remarkably, in the large $N_c, N_T$ limit, $M_W$ and $M_Q$ are fully $\rho$-independent quantities. However, we must keep in mind that, taking the Wilson-like terms of all the elementary fermions as $d=6$ operators, like we have done in this Appendix, may not be the best choice for phenomenology (see sect.~\ref{sec:SPCS}).

3) The explicit dependence of $m_q$, $m_\ell$ and $m_L$ upon the $\rho$ ratios ${\rho_q}/{\rho_Q}$, ${\rho_\ell}/{\rho_Q}$ and ${\rho_L}/{\rho_Q}$, respectively, may appear somewhat disturbing. The problem could be attenuated or even eliminated by conjecturing that some symmetry exists which, putting constraints on the $\rho$'s, restricts the variability range of their ratios. For instance, in the case where all Wilson-like terms are $d=6$ operators, an extreme but appealing situation would the one in which all the $\rho$'s are equal because of some underlying GUT symmetry. In this case the whole $\rho$ dependence would completely drop out from physical observables. 

4) We recall that the estimates in equations from~(\ref{MQUARK}) to~(\ref{MTELL}) represent fermion masses at asymptotic UV scales or better at the unification scale $\Lambda_{GUT}$, if the theory unifies. The issue of the running of masses from $\Lambda_{GUT}$ down to lower energies is discussed in sect.~\ref{sec:SPCS}.

\renewcommand{\thesection}{D} 
\section{The $WW$ binding energy in the non-relativistic approximation}  
\label{sec:APPD} 

In the non-relativistic approximation ($E_{bind}^2/4M_W^2 \ll 1$) the $WW$ interaction can be modelled by a particle of mass $m = M_W/2$ in the almost-square attractive potential well
\beqn
&&V(r)=-\frac{U_0}{\cosh^2(\omega r)}\, , \label{UR}\\
&&U_0 = c_U g_w^{\gamma_U} \Lambda_T \, ,\quad 
\omega = c_\omega g_w^{\gamma_\omega}\Lambda_T\, .
\label{PARAM}
\eeqn 
The particular $\cosh$-shape of the potential taken in eq.~(\ref{UR}) has been chosen only for practical reason, i.e.\  because for this potential the Schr\"odinger equation admits an exact analytic solution as shown in~\cite{LL}. At this stage we leave undetermined the $g_w$ dependence of $U_0$ and $\omega$. 

Concerning the magnitude of the coefficients $c_U$ and $c_\omega$, we remark that $c_\omega$ is very likely an ${\mbox O}(1)$ quantity since the $WW$ attractive interaction is supposed to stem from exchanges of Tera-meson or Tera-glueball resonances with O($\Lambda_T$) mass. Concerning $c_U$, under the reasonable assumption that the potential well depth should not exceed the Tera-strong scale, {\it viz.}\ $U_0 = c_U g_w^2 \Lambda_T \lesssim \Lambda_T$, we expect $c_U$ to be not too large since $g_w^2$ is only moderately small. The evaluation of $c_U$ is a delicate and important issue to be able to get a reliable determination of the non-relativistic estimate of $E_{bind}$. Naturally a more solid evaluation of the $WW$ binding energy would be offered by setting up and solving the Bethe--Salpeter equation for the $WW$ bound state. In sect.~\ref{sec:HIGGS} we give an estimate of $E_{bind}$ based on an approximate Bethe--Salpeter equation. 

Nevertheless, it is interesting to investigate the solution of the Schr\"odinger equation for the potential~(\ref{UR}). One finds the (negative) discrete energy eigenvalues~\cite{LL} 
\beqn
E_n=-\frac{\hbar^2\omega^2}{8m}\Big{[}-1+2n+\sqrt{1+\frac{8mU_0}{\hbar^2\omega^2}}\Big{]}^2\, , \quad m=\frac{M_W}{2} \, .
\label{EIGEN}
\eeqn
The allowed integers, $n$, are determined by the condition 
\beqn
n<\sigma \equiv \frac{1}{2}\Big{[}-1+\sqrt{1+\frac{8mU_0}{\hbar^2\omega^2}}\Big{]} \, .\label{QN}
\eeqn 
In our case, recalling~(\ref{PARAM}), we get ($\hbar=1$)
\beqn
\frac{8mU_0}{\hbar^2\omega^2} \quad \to \quad 
\frac{4M_W}{\Lambda_T} \frac{c_U }{c_\omega^2} g_w^{\gamma_U-2\gamma_\omega}\, ,\label{QUANT1}
\eeqn
and we require $\gamma_U-2\gamma_\omega\geq 0$. We can safely assume that the dimensionless quantity in~(\ref{QUANT1}) is small, either parametrically because one has in mind the limit $g_w\to 0$, or numerically because $c_U/c_\omega^2 \leq 1$, or finally because in our approach (see eq.~(\ref{MASSES})) $M_W= g_w c_w\Lambda_T$ with $c_w$ small (see sect.~\ref{sec:LF}) and $g_w<1$. From the definition of $\sigma$ given by eq.~(\ref{QN}) one can thus conclude
\beqn
\sigma \simeq \frac{M_W}{\Lambda_T} \frac{c_U }{c_\omega^2}g_w^{\gamma_U-2\gamma_\omega} < 1 \, .
\label{QUANT2}
\eeqn
In this regime only the bound state with $n=0$ exists and we find
\beqn
-E_{bind}=E_0= - \frac{ c_\omega^2 g_w^\kappa \Lambda_T^2}{4 M_W } 4\sigma^2 \simeq - \frac{c_U^2}{c_\omega^2} g_w^\kappa M_W \, , \qquad \kappa=2(\gamma_U-\gamma_\omega)\, .\label{EBNUM}
\eeqn
The calculation we have presented  shows that, given the parameters of the potential well ($\Lambda_T$) and the mass of the particle ($M_W$), there is only one bound state (at least ignoring spin and weak isospin indices) with a binding energy of the order of magnitude of $M_W$ itself times some power of $g_w$. 
As for the value of $\kappa$, our favourite choice for the $\gamma$-parameters in eq.~(\ref{PARAM}) is $\gamma_U=4\, ,\gamma_\omega=2$, yielding $\kappa=4$, as suggested by the Bethe--Salpeter-like arguments of  sect.~\ref{sec:ABT}.

\vspace{.2cm}
{\bf Acknowledgments -} We are indebted to R.\ Frezzotti for his interest in this work and for infinitely many comments and suggestions on the issues presented in this paper. We wish to thank R.\ Barbieri, M.\ Bochicchio, G.\ Martinelli, C.~T.\ Sachrajda, N.\ Tantalo, M.\ Testa and especially G.\ Veneziano for many useful discussions. We thank M.\ Garofalo for a careful reading of the manuscript and for illuminating correspondence.

\end{document}